# Analysis of Clumps in Saturn's F Ring from Voyager and Cassini

Robert S. French[a], Shannon K. Hicks[b], Mark R. Showalter[c], Adrienne K. Antonsen[d], Douglas R. Packard[e]

[a] SETI Institute, 189 Bernardo Ave., Mountain View, California 94043, United States; rfrench@seti.org
[b] SETI Institute, 189 Bernardo Ave., Mountain View, California 94043, United States; hicks.shannonk@gmail.com
[c] SETI Institute, 189 Bernardo Ave., Mountain View, California 94043, United States; mshowalter@seti.org
[d] SETI Institute, 189 Bernardo Ave., Mountain View, California 94043, United States; addieken@comcast.net
[e] SETI Institute, 189 Bernardo Ave., Mountain View, California 94043, United States; dpackar2@illinois.edu

## ABSTRACT

Saturn's F ring is subject to dynamic structural changes over short periods. Among the observed phenomena are diffuse extended bright clumps (ECs) ~ 3–40° in longitudinal extent. These ECs appear, evolve, and disappear over a span of days to months. ECs have been seen by the two Voyager spacecraft, the Cassini orbiter, and various ground- and space-based telescopes. Showalter (2004, *Icarus*, **171**, 356–371) analyzed all Voyager images of the F ring and found that there were 2–3 major and 20–40 minor ECs present in the ring at any given time. We expand upon these results by comparing the ECs seen by Voyager to those seen by Cassini in 2004–2010. We find that the number of minor ECs has stayed roughly constant and the ECs have similar distributions of angular width, absolute brightness, and semimajor axis. However, the common exceptionally bright ECs seen by Voyager are now exceedingly rare, with only two instances seen by Cassini in the six years, and they are now also much dimmer relative to the mean ring background. We hypothesize that these bright ECs are caused by the repeated impacts of small moonlets with the F ring core, and that these moonlets have decreased in number in the 25 years between missions.

# 1. INTRODUCTION

Saturn's F ring shows variations on many scales, both spatial and temporal. The features that have been identified go by a variety of names, have a range of properties, and illustrate a variety of dynamical processes. Examples include small embedded clumps or moonlets (Esposito et al., 2008; Meinke et al., 2012), jets which shear to kinematic spirals (Charnoz, 2009; Charnoz et al., 2005; Murray et al., 2008), streamer-channels (Murray et al., 2005; Porco et al., 2005), and mini-jets (Attree et al., 2012). In this paper, we focus on a phenomenon we refer to as "extended bright clumps", defined as localized bright regions ~ 3–40° in longitudinal extent. This particular class of features in the ring can be identified using purely objective criteria, enabling us to compare the Voyager and Cassini data sets in a rigorous, statistical manner. Our analysis specifically excludes the very localized clumps or moonlets ($\ll 3°$ in extent), which have been studied by others (Beurle et al., 2010; Esposito et al., 2008; Hedman et al., 2011; Meinke et al., 2012).

Extended clumps, which we call "ECs", have been seen by Pioneer 11 (Gehrels et al., 1980), Voyager 1 and 2 (Smith et al., 1982, 1981), the Hubble Space Telescope (Bosh and Rivkin, 1996; McGhee et al., 2001; Nicholson et al., 1996; Poulet et al., 2000b), ground-based telescopes (Charnoz et al., 2001; Poulet et al., 2000a; Roddier et al., 2000), and the Cassini Orbiter (Porco et al., 2005). When an EC is detected multiple times we refer to it as a multiply-detected clump or "MDC"; MDCs are distinct only in the sense that we can measure their changes over time, analyze their orbits, and place bounds on their lifetimes. Showalter (2004) (henceforth S04) examined images from Voyager 1 (1980) and Voyager 2 (1981) and found that there were 2–3 major and 20–40 identifiable smaller ECs present at any given time. He also analyzed the orbits and lifetimes of 34 MDCs. The MDCs had a distribution of mean motions relative to the F ring's core that indicated they had semimajor axes spread over ~ 100 km and had relatively short lifetimes, with no MDCs surviving the nine-month gap between the two Voyager visits.

In this work we conduct an analysis similar to S04's, but using six years of images from Cassini. The longer temporal baseline and higher resolution images allow us to identify and analyze a variety of ECs, and in some cases watch MDCs being created or destroyed. We compare our results with those from S04 to identify changes in the F ring during the 24 years between missions, and we use our higher-resolution images to analyze the physical characteristics and morphology of the ECs. We further speculate on the roles that small embedded moonlets and the "shepherd" moon Prometheus play in EC formation and destruction.

In Section 0 we describe our methodology for image selection and processing for both Cassini and Voyager data in the detection and tracking of ECs. In Section 3 we examine the physical characteristics of ECs including their longitudinal extent and brightness. In Section 4 we analyze the orbits of the MDCs relative to the F ring core and discuss their creation, destruction, and lifetimes. We also examine how MDCs change over time and identify several unusual MDCs. In Section 5 we investigate possible causes for EC formation, including the role of embedded moonlets and Prometheus. Finally, in Section 6 we discuss our conclusions.

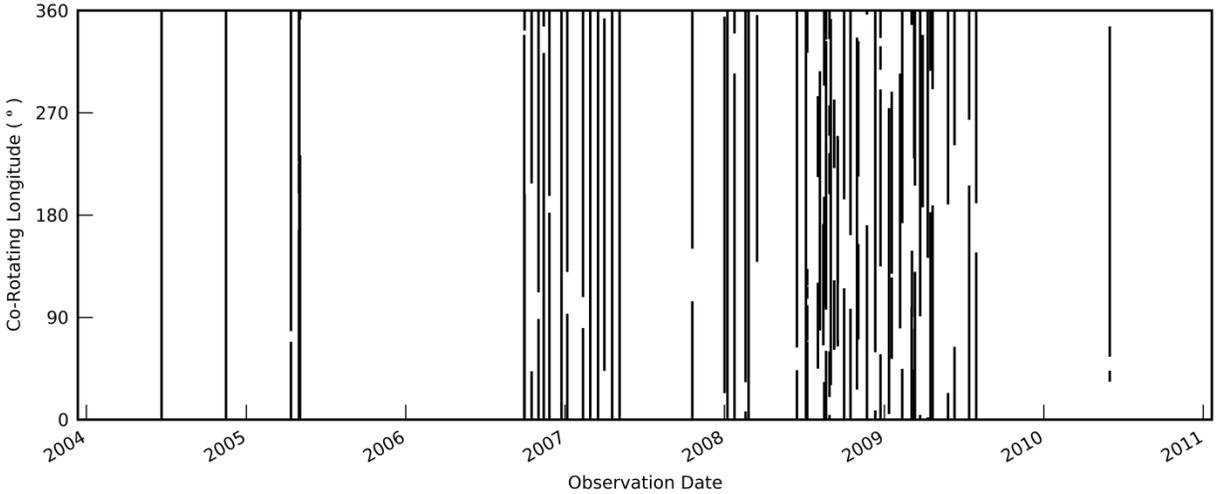

**Figure 1:** Longitudinal coverage of Cassini ISS observations of the F ring used in this study. Each vertical line represents one profile and is drawn where coverage is available. Cassini observations are not distributed evenly in time and most do not provide complete 360° coverage.

## 2. METHODOLOGY

*2.1. Cassini ring profiles*

The Cassini Imaging Science Subsystem (ISS) includes two cameras, the Narrow Angle Camera (NAC) with a 0.35°× 0.35° field of view and the Wide Angle Camera (WAC) with a field of view 10 times larger in each dimension (Porco et al., 2004). We reviewed all images of the rings taken through the clear filters (CL1 and CL2) obtained from the beginning of the Saturn tour through mid-2010 to find those that included the F ring. As the goal was to produce longitudinal profiles of ring brightness for further analysis, we chose sequences of high-quality images taken in rapid succession that, together, covered the majority of the F ring (Table 1). Although some of the images were taken with the WAC, most of the images we have used were taken by the NAC. The NAC images were often taken in groups as scheduled "movies", image sequences consisting of a few dozen to over 1,000 images taken within a period of 10 to 15 hours. During these movies, Cassini stared at a nearly-fixed inertial longitude, most often the ring ansa, obtaining partial or complete longitudinal coverage of the F ring as it rotated beneath the spacecraft. Regardless of the manner in which the images were taken, we combined image sequences together to form a series of reprojected mosaics, as described below.

In total, we used 65 image sequences or "observations", each labeled with a Cassini observation identifier, consisting of nearly 9,000 calibrated images retrieved from the Rings Node of NASA's Planetary Data System. Nine of these sequences contained complete 360° coverage and 33 contained more than 80% coverage. The available observations were inhomogeneously distributed in time, giving excellent coverage during some time periods but no coverage at all during others (Figure 1).

We used the SPICE kernels (Acton, 1996) to determine the pointing direction and field of view of the appropriate ISS camera for each image. This pointing has small inaccuracies, so a combination of automated detection of the F ring core (defined as the brightest string of pixels with the correct orientation and curvature) and manual adjustment was used to determine the exact location of the core in each image. We modeled the F ring using the orbital elements from Albers et al. (2012) as a non-inclined, freely-precessing ellipse, and assumed that the brightest portion of the core falls at a constant semimajor axis. Any small variation in the location or eccentricity (Albers et al., 2012; Bosh et al., 2002; Cooper et al., 2013) of the core will not affect our fundamental results.

We reprojected the F ring in each image into a linear space where one dimension is distance from the F ring core and the other is the mean orbital longitude in equal steps from 0° to 360°. The use of mean longitude allows us to easily compute mean motion directly from two or more observations by dividing the difference in mean longitude by the intervening time. Because we are only interested in motion relative to the core, we adjusted the mean longitude to be relative to a reference frame co-rotating with the F ring, assuming a mean motion of 581.979°/day and an origin at the intersection of the ascending node of Saturn's ring plane and Earth's equator at the epoch January 1, 2007 00:00:00 UTC. For simplicity and compatibility with previous works, we refer to this adjusted mean longitude as "co-rotating longitude" throughout the rest of this paper.

We sampled each image at increments of 5 km in radius and 0.02° in co-rotating longitude. As one degree of longitude is ~ 2447 km of arc length at the semi-major axis of the F ring core, our longitudinal samples are ~ 49 km in length. To create a final mosaic from each observation, we combined the various reprojected images. Where images overlapped, for each co-rotating longitude we chose the radial strip from the image that had the most complete radial coverage; when there was a tie, we chose the image with the finest radial resolution.

Each mosaic was then analyzed for anomalous pixels (including bad pixels, transmission errors, and the presence of Prometheus or Pandora) and these pixels were masked from future use. Finally, we subtracted from each radial slice a quadratic background model that matched the empty space both interior to and exterior to the F ring (cf. French et al. 2012). Pixel values in calibrated images are given as $I/F$, where intensity $I$ is measured relative to the incident solar flux density $\pi F$. This quantity is dimensionless and scaled such that $I/F = 1$ for a perfectly diffusing, flat Lambertian surface illuminated from normal incidence. Every radial slice of a mosaic was integrated to produce its "equivalent width", defined as:

$$W = \int I/F(a)da, \qquad (1)$$

where $a$ is the radial distance from Saturn. We refer to the collection of equivalent widths for a particular mosaic as its "profile", a representation of intensity with co-rotating longitude (Figure 2). This method of computing equivalent width allows us to use lower-resolution images, permitting the effective comparison of the Voyager and Cassini data sets. However, it removes the details of the radial structure of the ring, such as the presence of mini-jets, streamer-channels, and additional ring strands.

## 2.2. *Voyager ring profiles*

We took our Voyager data directly from S04, who analyzed nearly every image of the F ring taken by the two Voyager spacecraft: 869 images from Voyager 1 taken over ~ 55 days and 618 images from Voyager 2 taken over ~ 35 days. The processing of Voyager images by S04 followed the same overall strategy that we used for Cassini images. However, many of the images were unresolved, and the wide Voyager PSF may have caused ring brightness to spread into the adjacent background regions, resulting in inaccurate photometry. We discuss how we have compensated for this source of error in Section 2.3.

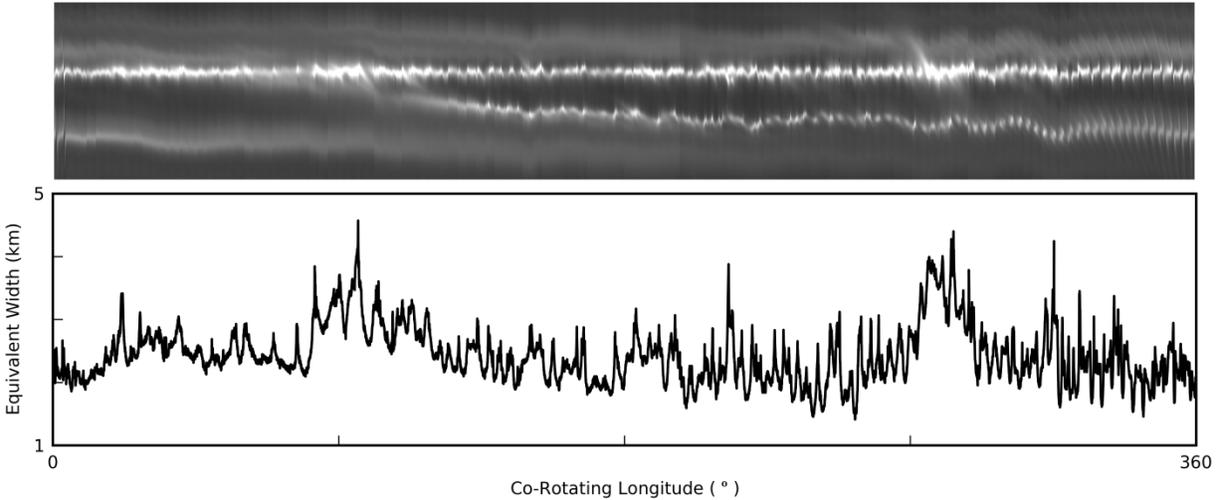

**Figure 2: The mosaic from Cassini observation ISS_059RF_FMOVIE002_VIMS and its associated equivalent width profile. The radial limits of the mosaic are 139,800 (bottom) to 140,500 km (top). The mosaic has been contrast enhanced for better visibility. Occasional, small radial alignment errors during the mosaicking process do not affect the derived profile.**

Because each image only contained a portion of the ring, and images were not generally taken in rapid succession, the Voyager data set has a fundamentally different character than our Cassini data. With Cassini we have 33 nearly-complete (and another 32 less-complete) "movies", each taken within a ~ 15-hour period but spread over six years, yielding gaps of weeks to months when the F ring was not observed. In contrast, with Voyager we have nearly 1500 images, each of which shows only part of the ring, all taken within 1–2 months. Voyager imaged any particular portion of the F ring every 5–15 days, giving much better time resolution. However, the total coverage by both Voyager spacecraft was only ~ 90 days, giving an isolated snapshot of the ring's characteristics compared to the six years provided so far by Cassini.

S04 produced profiles, similar to the ones we produced from Cassini data but with a longitudinal resolution of 0.5°, for each of the inbound and outbound phases of the two Voyager spacecraft. These four profiles merge together several weeks of data. Even though ECs may have shifted slightly during the creation of each profile, the profiles are nevertheless useful for measuring the overall characteristics of the F ring and the physical characteristics (number, angular width, and brightness) of the ECs, which were not computed by S04.

*2.3. The F ring phase curve*

It is well known that the brightness of the dusty F ring is a strong function of viewing geometry and phase angle (French et al., 2012; Showalter et al., 1992). To better compare profiles and clumps across observations, we normalized both Cassini and Voyager profiles using two different methods.

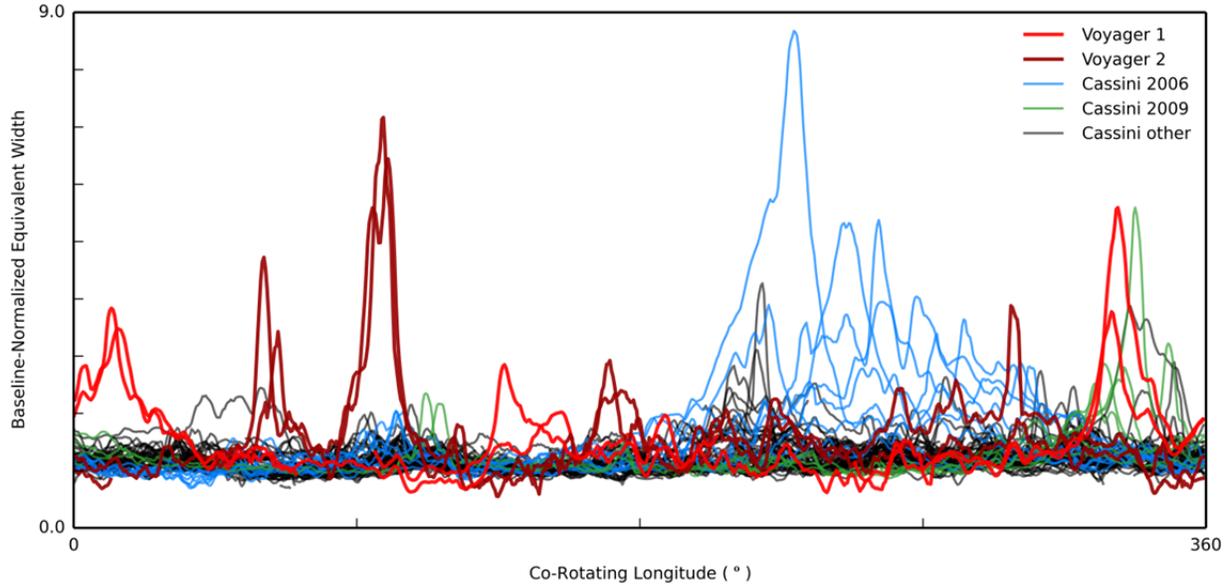

**Figure 3: Comparison of baseline-normalized profiles from Voyager and Cassini.** All profiles are normalized to the 15th percentile value representing the non-clump brightness of the ring. The Voyager profiles are V1I, V1O, V2I, and V2O from S04. All Cassini profiles are included. The profiles from December 23, 2006 to May 5, 2007 (showing C19/2006) and the profiles from March 11, 2009 to June 10, 2009 (showing C54/2009) are marked separately.

In the first method, we normalized each profile relative to its non-clump baseline brightness. We found that on average no more than ~ 75%, and never more than ~ 85%, of the ring is occupied by clumps at any given time. Thus, for each profile we chose the 15th percentile value as representative of this background brightness and divided each equivalent width in the profile by this value. The resulting dimensionless "baseline-normalized equivalent width" illustrates how much brighter each clump is compared to the background of the ring at the same time. All baseline-normalized Cassini and Voyager profiles are shown in Figure 3.

The second method of normalization allows us to adjust the absolute brightness of the ring to that which would be seen from an observer looking normal to the ring plane at a phase angle $\alpha$ of 0°. To enable this normalization, we first computed the phase curve of the F ring using the (more numerous) Cassini profiles, employing a methodology similar to that used by French et al. (2012) as described below.

The F ring has sufficient optical depth that obscuration and shadowing are important. To compute the brightness that would be seen normal to the ring plane, we multiplied each equivalent width $W$ by $\mu \equiv |\cos(e)|$, where $e$ is the emission angle of that particular measurement measured relative to the ring plane normal. The resulting value is called the *normal equivalent width*. We then adjusted this value to account for shadowing and single scattering by multiplying by one of two adjustment factors

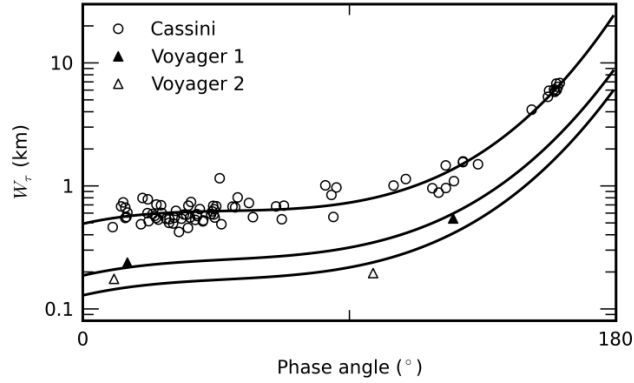

**Figure 4:** Cassini and Voyager 15$^{th}$ percentile $\tau$-adjusted normal equivalent widths and associated phase curves. Cassini measurements with $|B_0| > 3°$ and Voyager measurements from the four available profiles are included. The associated 3rd-order fit polynomials are shown; in each case the shape of the fit curve is constrained to be the same as the Cassini curve. The bottom curve is the fit to Voyager 2, the middle curve is the fit to Voyager 1, and the top curve is the fit to Cassini.

$$z_R(\tau,\mu,\mu_0) = \frac{\tau(\mu+\mu_0)}{\mu\mu_0(1-\exp(-\tau(1/\mu+1/\mu_0)))} \quad (2)$$

or

$$z_T(\tau,\mu,\mu_0) = \frac{\tau(\mu-\mu_0)}{\mu\mu_0(\exp(-\tau/\mu)-\exp(-\tau/\mu_0))}, \quad (3)$$

depending on whether the ring was seen in reflected ($Z_R$) or transmitted ($Z_T$) light, yielding the "$\tau$-adjusted normal equivalent width" or $W_\tau$ (cf. French et al. 2012). In these formulas $\mu_0 \equiv |\cos(i)|$, where $i$ is the incidence angle measured relative to the ring plane normal, and $\tau$ is the equivalent optical depth of the ring (cf. French et al. 2012). Unfortunately, the approximations we use for shadowing and single-scattering break down when the ring opening angle to the Sun $B_0 = 90°-i$ is small, which occurred near the Saturnian equinox in 2009. As a result, we eliminated any observations with $|B_0| < 3°$ from our phase curve, leaving 50 profiles.

Because the clumps cause drastic variations in the brightness of the F ring with longitude, we computed the phase curve based on the background, non-clump portions of the ring, using the same 15$^{th}$ percentile value described above in the first normalization method. Our use of this baseline also obviated the need to perform the additional modeling of changing clump brightness done by French et al. (2012).

The phase curve is based on the mean phase angle for each profile. However, some profiles were created from images that span fairly wide ranges in phase angle. In these cases, we broke the profile into multiple, smaller profiles, where each had a more limited range of phase angles. We modeled the phase curve as a simple cubic polynomial fit to $\log_{10}(W_\tau(\alpha))$, where each datum was weighted by the longitudinal coverage of the ring available in that profile. We found $\tau = 0.035$ by iteration; it is the value that minimized the scatter of the observations about the phase curve. The resulting phase curve (with $\alpha$ given in degrees) is $\log_{10}(P(\alpha)) = a\,\alpha^3 + b\,\alpha^2 + c\,\alpha + d$, with $a = 6.099\times10^{-7}$, $b = -8.813\times10^{-5}$, $c = 5.517\times10^{-3}$, and $d = -0.3296$ (Figure 4). This result is consistent with the value $\tau = 0.033 \pm 0.008$ found by French et al. (2012) using substantially fewer measurements.

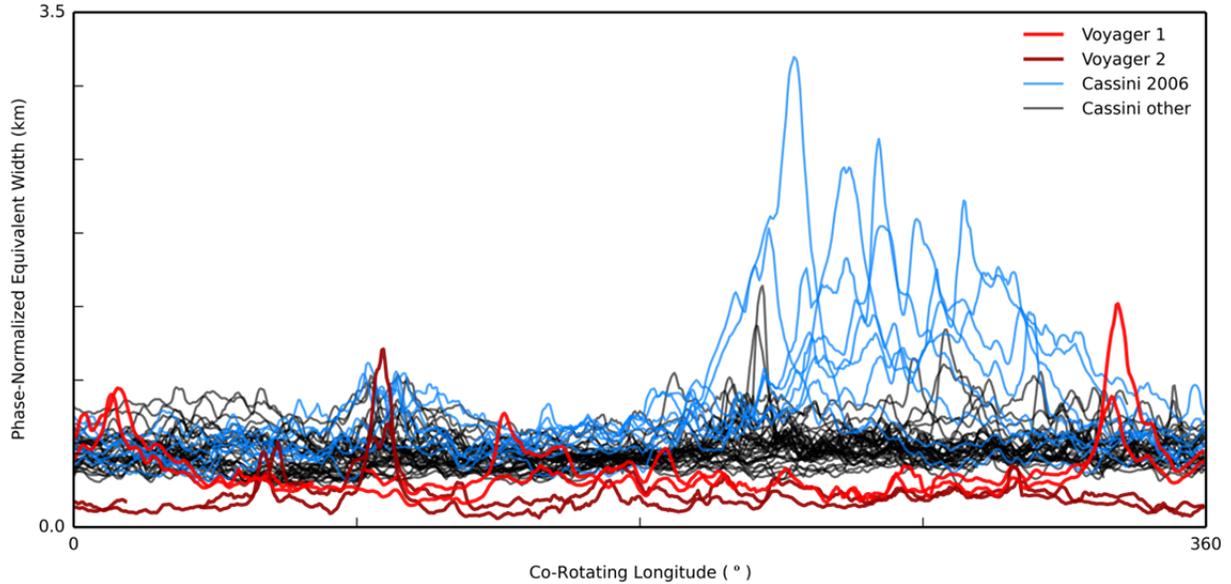

**Figure 5:** Comparison of phase-normalized profiles from Voyager and Cassini. All profiles are normalized to $e = 0°$ and $\alpha = 0°$. The Voyager profiles are V1I, V1O, V2I, and V2O from S04. All Cassini profiles with $|B_0| > 3°$ are included. The profiles from December 23, 2006 to May 5, 2007 (showing C19/2006) are marked separately. The profiles from February 5, 2009 to July 30, 2009 (which include those showing C54/2009) are not shown due to the opening angle limitation.

Given this phase curve, we corrected for the effect of viewing from a non-zero phase angle by multiplying each $W_\tau$ by $P(0)/P(\alpha)$, yielding the "phase-normalized equivalent width." We assumed that the phase curve and optical depth are the same for the Cassini and Voyager observations, and performed the same adjustment to the four Voyager profiles using the Cassini-derived values. To compensate for any error in the Voyager photometry, we compared the photometrically-accurate Voyager data from Showalter et al. (1992), which provided the mean brightness of the ring at a variety of phase angles, to the mean brightness of the four profiles from Showalter (2004). We found that S04's Voyager 1 measurements needed to be increased by 34% and S04's Voyager 2 measurements needed to be decreased by 3%, and we applied this adjustment for all further analysis. All phase-normalized Cassini and Voyager profiles are shown in Figure 5.

*2.4. Detection of extended clumps using wavelets*

For convenience and efficiency, we used wavelets (Addison, 2002; Torrence and Compo, 1998) to detect extended clumps in both Cassini and Voyager profiles. The continuous wavelet transform, defined as

$$X_w(a,b) = \frac{1}{\sqrt{a}} \int_{-\infty}^{\infty} x(t) \psi^* \left( \frac{t-b}{a} \right) dt , \quad (4)$$

takes a square-integrable data series $x(t)$ (in our case, a ring profile) and correlates it with different scales $a$ of a "mother wavelet" $\psi$ at each location $b$ in the data. The result is a two-dimensional representation of the data called a scalogram, where each value in the scalogram gives the correlation of a particular scale of the mother wavelet at a particular location in the data (Figure 6).

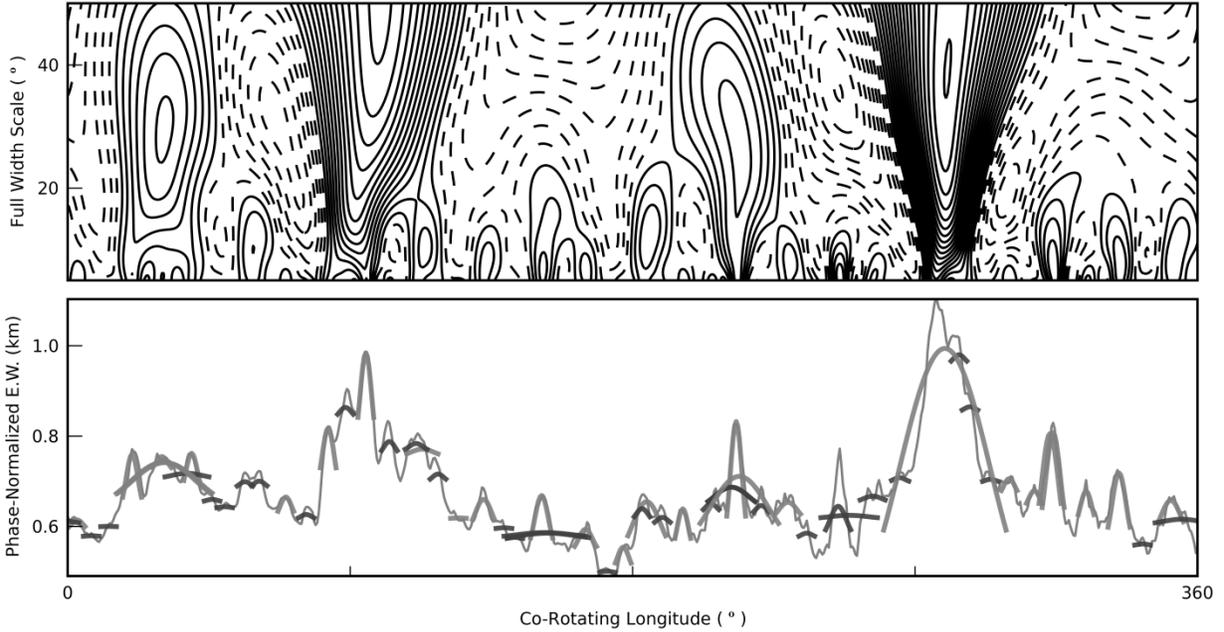

**Figure 6: Sample scalogram of Cassini observation ISS_059RF_FMOVIE002_VIMS using the SDG mother wavelet (top) and clump candidates found using the SDG (light gray) and FDG (dark gray) wavelets (bottom). The full width scale (top panel, vertical axis) is twice the wavelet scale *a* in Equation (4). Local maxima in the scalogram indicate clump locations and sizes. Contours of negative values are shown with dashed lines.**

To be used with the continuous wavelet transform, a mother wavelet must have a mean of zero, a square norm of one, and must satisfy an admissibility criterion; see Addison (2002) for full details. This places significant constraints on the shapes of wavelets that may be used. S04 modeled clumps with a simple Gaussian, which does not meet the criteria for a valid mother wavelet. However, the 2nd derivative of the Gaussian ("SDG") and the 4th derivative of the Gaussian ("FDG") functions have very similar shapes and can be used as mother wavelets (Figure 7):

$$\psi_{SDG}(x) = \left(1 - x^2\right)\exp(-x^2/2) \tag{5}$$

$$\psi_{FDG}(x) = \left(3 - 6x^2 + x^4\right)\exp(-x^2/2) \tag{6}$$

To detect ECs in a profile, we first passed the data through a low-pass filter that took the running mean over a 3° window. This removed local high-frequency variations that are unrelated to clump detection and also helped smooth out some of the ~ 3.2° features caused by the F ring's interactions with Prometheus (see Section 5). We then computed the continuous wavelet transform using both the SDG and FDG wavelets over scales of 3.5° to 40°, producing two scalograms. We chose 3.5° as the lower cutoff to prevent the detection of structures caused by the passage of Prometheus and 40° as the upper cutoff to prevent large random variations from producing false detections. Within each scalogram, every local maximum indicates the location and scale of a potential clump. We took the union of the sets of clumps found using the two wavelets. When a clump was found by both wavelets, we fitted the wavelet shapes to the clump

- 9 -

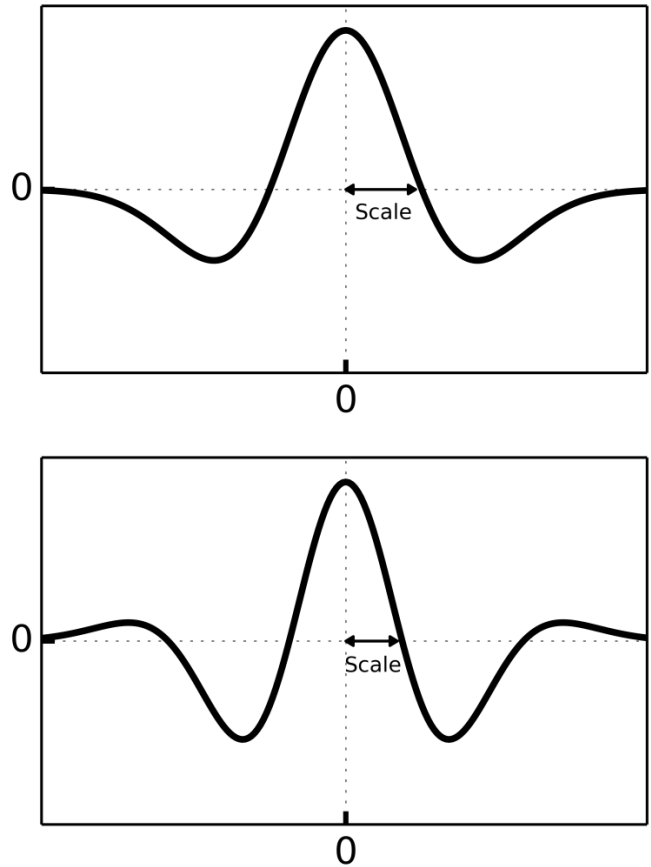

**Figure 7: Second derivative of Gaussian (SDG, top) and fourth derivative of Gaussian (FDG, bottom) mother wavelets. No axis scales are shown as wavelets are scale invariant.**

and took the version that has the smallest residual. In almost all cases, the SDG was sufficient and preferable for clump detection. However, we encountered rare cases when the SDG missed particular clumps, which were then found successfully by the FDG wavelet, so we found it useful to include the FDG wavelet in our processing. Note that in some cases multiple clumps can refer to the same areas of a profile, as clumps or clusters of clumps can be detected at multiple scales (Figure 6).

A large, anomalous bright clump visible for several months after December 2006 (French et al., 2012; Murray et al., 2008), henceforth called C19/2006, and a similar but less dramatic clump in early 2009 (C54/2009) were not successfully found using our wavelet technique and were added by hand to the wavelet results. We removed the 40° limit on angular width when processing these clumps. As there were no clumps this wide in the Voyager data, this difference in processing did not affect our statistics.

As noted previously, the use of equivalent width removes the details of the radial structure of the ring. As such, it is likely that some ECs detected by the wavelet analysis were in secondary strands rather than in the main core. However, such ECs would have been detected equally in the resolved Cassini and unresolved Voyager images, making any comparison of physical characteristics consistent between the two missions.

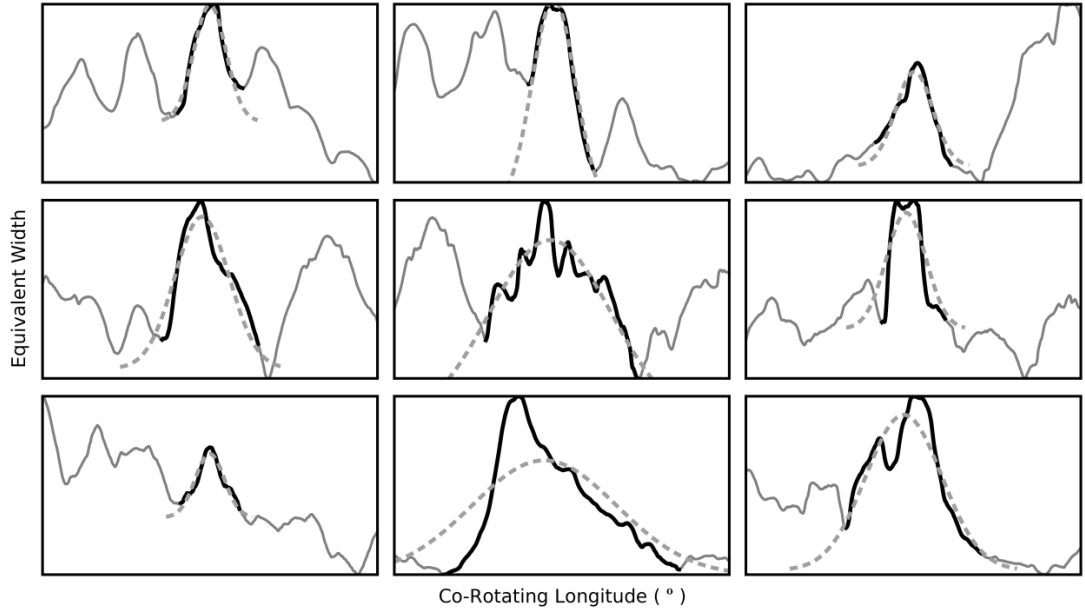

**Figure 8: Sample clumps (black) with best-fit Gaussian curve (dotted). Each plot shows a 30° section of the F ring. The boundaries of the clump are determined from the minima within 2 σ. The Gaussian is drawn to 3 σ on each side for illustration. The vertical axis is not the same scale in each plot. From left to right, then top to bottom, the clumps are the earliest observations of C2, C6, C9, C13, C15, C30, C31, C35, and C39.**

The wavelet fit can be used as a first approximation of the center and angular width of a clump, but the negative extent of the SDG and FDG wavelets, as well as the asymmetric shapes of many ECs, make these values somewhat unreliable. We thus next refined our determination of the center and angular width by fitting a Gaussian plus a constant to the data in the general vicinity of the clump detected by the wavelet. We allowed the Gaussian to extend a different distance on each side of the clump center to accommodate asymmetric clumps. We define the location of the clump to be the center of the Gaussian. We define the "angular width" of the clump to be the distance between the minima of the profile data on either side of the center within two standard deviations (Figure 8). We note that our angular width measurements of clumps are at least partially dependent on the shapes, as non-Gaussian shapes will be fit as closely as possible but the 2 σ size limit will depend on the width of the Gaussian chosen. We further define the "integrated brightness" of a clump as the integrated baseline- or phase-normalized equivalent width over the angular extent of the clump after a linear background has been subtracted:

$$B = \int_{l_0}^{l_1} W_l \, dl - (W_{l_0} + W_{l_1})/2, \qquad (7)$$

where $l$ represents co-rotating longitude and $l_0$ and $l_1$ are the starting and ending longitudes. As our phase-normalized equivalent width is in units of km, integrated brightness is in units of km-degrees; likewise baseline-normalized integrated brightness is in units of degrees. Finally, we define the "peak brightness" of a clump to be the maximum difference between the baseline- or phase-normalized equivalent width of any longitude in the clump and its associated linear background. Phase-normalized peak brightness is in units of km while baseline-normalized peak brightness is dimensionless.

Based on these fits, we eliminated clumps that were outside of our valid angular width range, were excessively asymmetric, or were too dim to be of interest. In total we found 2016 Cassini ECs and 116 Voyager ECs. Table 1 lists the number of ECs found in each Cassini observation.

*2.5. Multiply-detected clumps*

Where possible, we tracked individual Cassini ECs across multiple profiles, producing a list of multiply-detected clumps (MDCs). We automatically detected the presence of the same EC in multiple (two or more) observations by looking for a near-linear change in co-rotating longitude with time (Figure 9). We limited our detections to changes of ~ 0.7°/day or less, corresponding to a ~ 110 km range of semimajor axes. All detections of MDCs were visually confirmed and only MDCs that were present solely in the main F ring core were included.
We identified a total of 58 MDCs in the Cassini data (Table 2), each tracked through two to eight observations. Given the sparse nature of the Cassini observations, the detection of MDCs is limited by temporal and longitudinal coverage. We do not claim to have detected every possible EC appearing in multiple profiles, but instead have produced a set of MDCs containing no false positive identifications, suitable for statistical analysis.

We took the list of Voyager MDCs directly from S04, as he derived mean motions from a large number of detailed images rather than from the small number of complete profiles available to us. Seven of the 34 clumps tracked by S04 (their 1I, 1L, 1L′, 1M, 1N, 2C′, and 2I) were eliminated because they were so small that the filtering process erased them from the profiles, preventing the determination of their physical characteristics, and clump 1D was eliminated because S04 did not provide a mean motion measurement. This yielded 26 Voyager MDCs. However, S04 presented the results from Voyager assuming a semimajor axis of 140,185 km for the F ring core (Synnott et al., 1983), resulting in a systematic bias to the relative mean motions. We corrected for this bias by adding 0.306 °/day to the results of S04.

## 3. PHYSICAL CHARACTERISTICS OF EXTENDED CLUMPS

Our clump-finding process allows us to compute the angular width, integrated brightness, and peak brightness of extended clumps in an objective manner. Because the Cassini and Voyager profiles have different resolutions, we first down-sampled the Cassini profiles by taking the mean of the data in 0.5° increments and reran the clump-finding procedure described in Section 2.3 to provide an accurate comparison.

The continuous wavelet transform can find clumps at multiple scales at the same approximate longitude, and often finds smaller clumps that compose a larger clump. For the purposes of comparing physical characteristics, we have chosen to examine only the largest clumps because these large clumps tend to behave as single bodies; the smaller clumps that compose them are simply internal detail without separate behaviors. Thus we eliminated overlapping clumps by choosing the largest clump that fully enclosed any smaller clumps, leaving at most one clump at any given longitude.

We then removed the clumps related to C19/2006 and C54/2009 from our statistics as they are rare events. However, we retained the brightest Voyager clumps as these clumps were common. Finally, because we are unable to compute phase-normalized equivalent widths for observations near the equinox, we removed any clumps with $|B_0| < 3°$; unfortunately this also includes C54/2009, preventing further quantitative analysis, as discussed below. Although we were not able to analyze the brightness of these clumps quantitatively, visual inspection indicated that, with the exception of C54/2009, they are qualitatively similar in distribution to the remaining clumps. Thus, we do not expect their removal to affect our statistics. In the end, we were left with 881 ECs for Cassini and 81 ECs for Voyager.

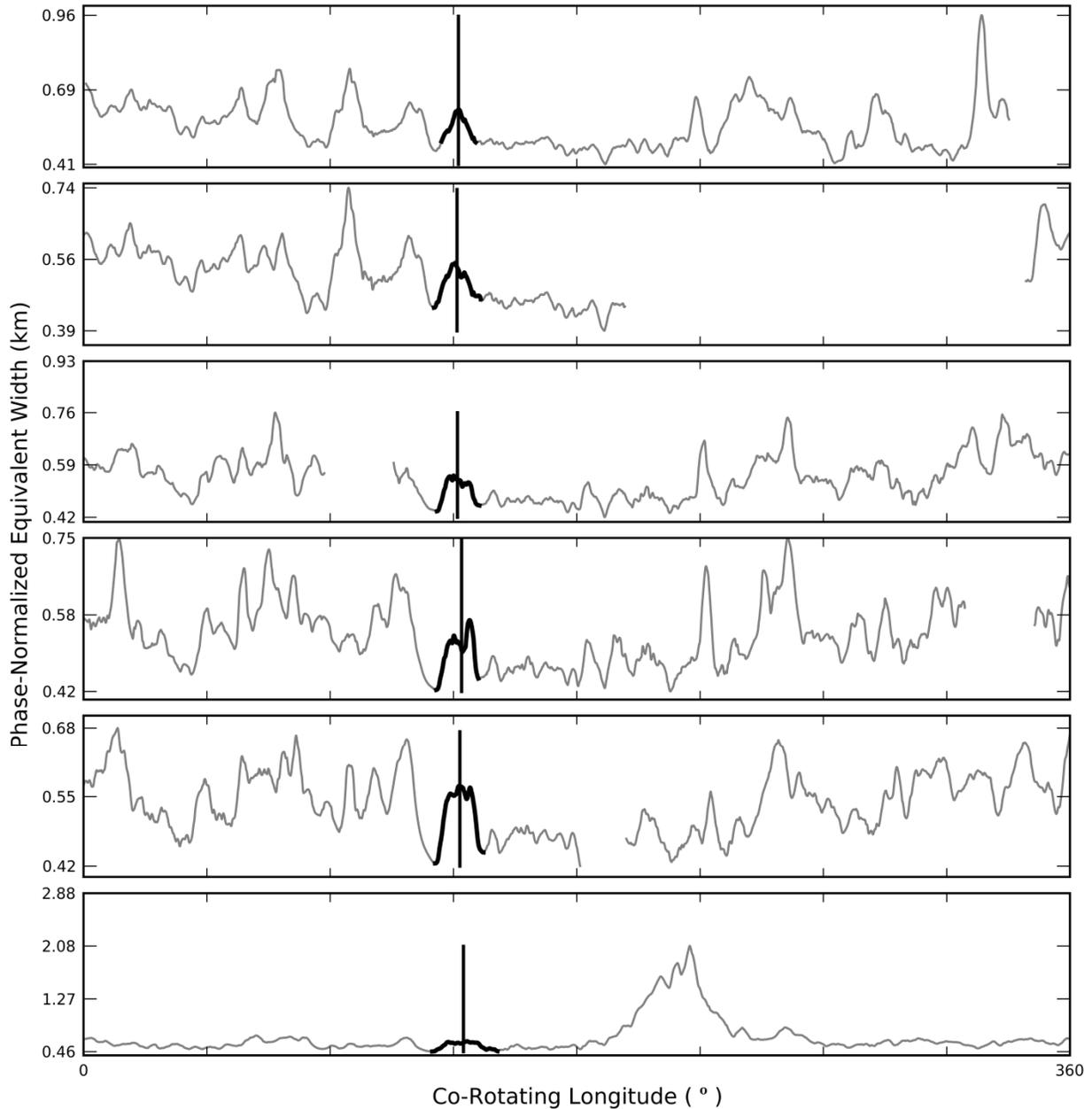

**Figure 9: Progression of clump C10 through six observations (from the top, ISS_029RF_FMOVIE001_VIMS, ISS_029RF_FMOVIE002_VIMS, ISS_031RF_FMOVIE001_VIMS, ISS_032RF_FMOVIE001_VIMS, ISS_033RF_FMOVIE001_VIMS, and ISS_036RF_FMOVIE001_VIMS). The center of the Gaussian fit to the clump is shown with a vertical line. The final panel also shows the appearance of C19/2006.**

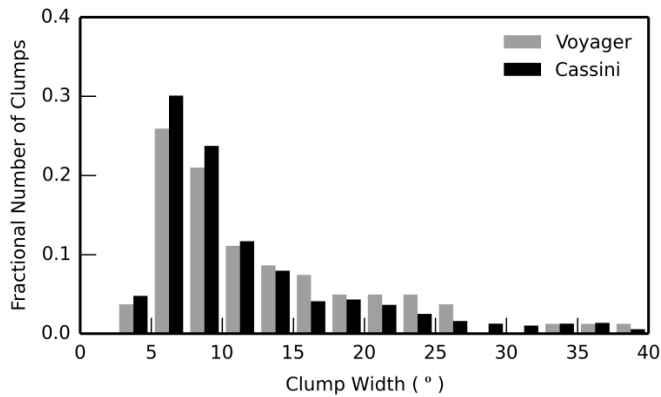

**Figure 10: Distribution of angular widths for ECs from Voyager and Cassini observations with $|B_0| > 3°$. ECs composing the Cassini C19/2006 and C54/2009 MDCs are not included. The majority of clump widths range from ~ 4° to ~ 10° and both distributions are roughly similar, with a cut-off at the low end because we ignore clumps smaller than 3.5°, and a fairly long tail.**

The physical characteristics of the Voyager and Cassini ECs are summarized in Table 3. The angular widths (Figure 10), phase-normalized integrated brightnesses (Figure 11), and phase-normalized peak brightnesses (Figure 12) have roughly the same ranges and means, and a Kolmogorov-Smirnov test fails to show that the distributions are different. However, the histograms of Voyager clump brightness show a distinct excess of points at the right (brighter) end of the histogram, associated with the bright clumps that were common in the Voyager profiles but nearly absent from the Cassini data. It should be noted, however, that the large bright 2006 and 2009 events were excluded from this histogram; this is discussed further below.

We find a different story when using the baseline-normalized profiles. The baseline-normalized integrated brightnesses (Figure 13) and baseline-normalized peak brightnesses (Figure 14) show that the Voyager clumps were, overall, significantly brighter than the Cassini clumps relative to their own baselines. As the absolute phase-normalized brightness measurements were similar, this difference is clearly caused primarily by changes in the brightness of the mean background ring (French et al., 2012), although the excess number of very bright clumps also plays a role. In summary, the population of "typical" clumps is very similar in the Voyager and Cassini data sets, although clumps tend to be more visible in the Voyager data set just because the ring's baseline brightness is lower. Nevertheless, the number of bright clumps seems to have changed from Voyager to Cassini, because in the Voyager era 2–3 bright clumps were typically seen at any given time, whereas in the Cassini data only two very bright clumps, each lasting ≲ 3–6 months, have occurred within a 6-year period.

We investigated the statistical significance of this apparent change by examining the null hypothesis that the mean number of very bright ECs present at any given time, which we call $\lambda$, was the same during the Voyager and Cassini missions. As discussed in Section 4.2, we assume that the creation of ECs is a Poisson process wherein the probability of seeing $n$ bright ECs at a time is $P_\lambda(n) = \lambda^n e^{-\lambda} / n!$. Since no ECs survived between the two Voyager missions, the probability that each Voyager observed two or more bright ECs is

$$V(\lambda) = [1 - P_\lambda(0) - P_\lambda(1)]^2. \tag{8}$$

Likewise, the probability that Cassini observed two or fewer bright ECs is

$$C(\lambda, N) = P_\lambda(0)^N + NP_\lambda(0)^{N-1}P_\lambda(1) + \left[\frac{N(N-1)}{2}P_\lambda(0)^{N-2}P_\lambda(1)^2 + NP_\lambda(0)^{N-1}P_\lambda(2)\right], \quad (9)$$

where $N$ is the number of distinct time periods such that a bright EC is unlikely to survive from one period to the next. The exact value for $N$ is difficult to determine. A lower bound based on a maximum lifetime of ~ 9 months (Section 4.2) and the available Cassini coverage (Table 1, Figure 1) is $N \cong 5$, while an upper bound based on a maximum lifetime of ~ 3 months is $N \cong 10$.

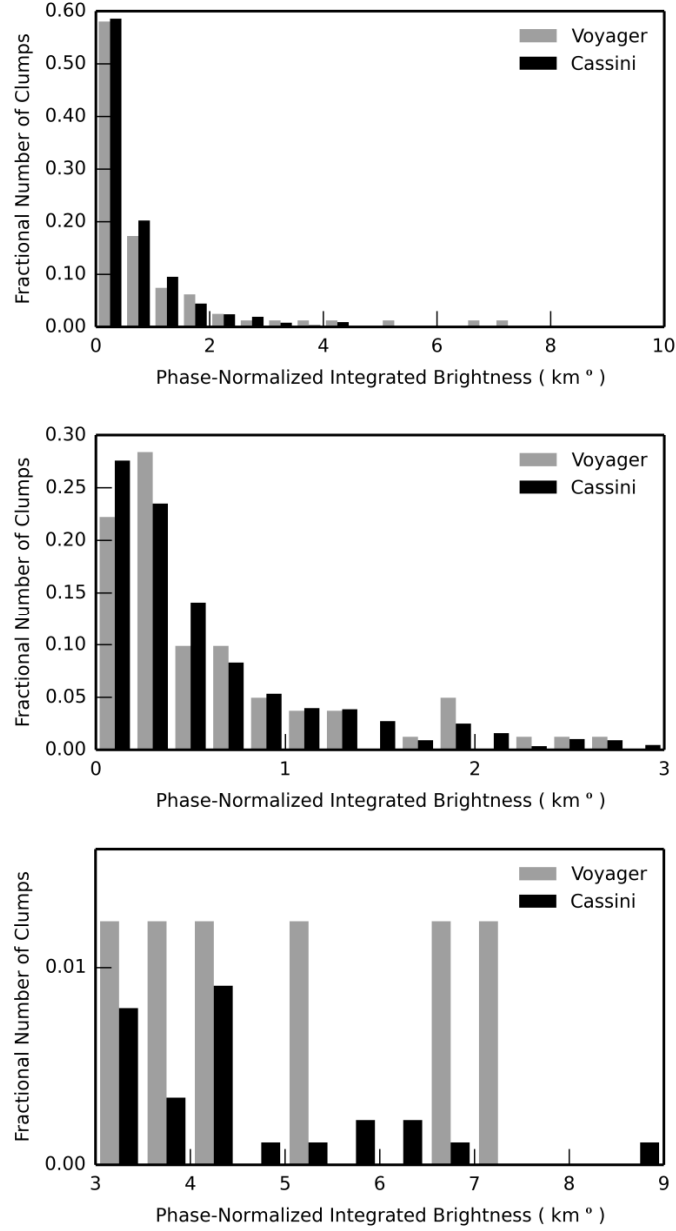

**Figure 11: Distribution of phase-normalized integrated brightness for ECs from Voyager and Cassini observations with $|B_0| > 3°$. ECs composing the Cassini C19/2006 and C54/2009 MDCs are not included. The top histogram shows all of the clumps, while the middle and bottom histograms have been zoomed in to show only data on the dimmest and brightest clumps, respectively. The excess of very bright clumps seen by Voyager is apparent.**

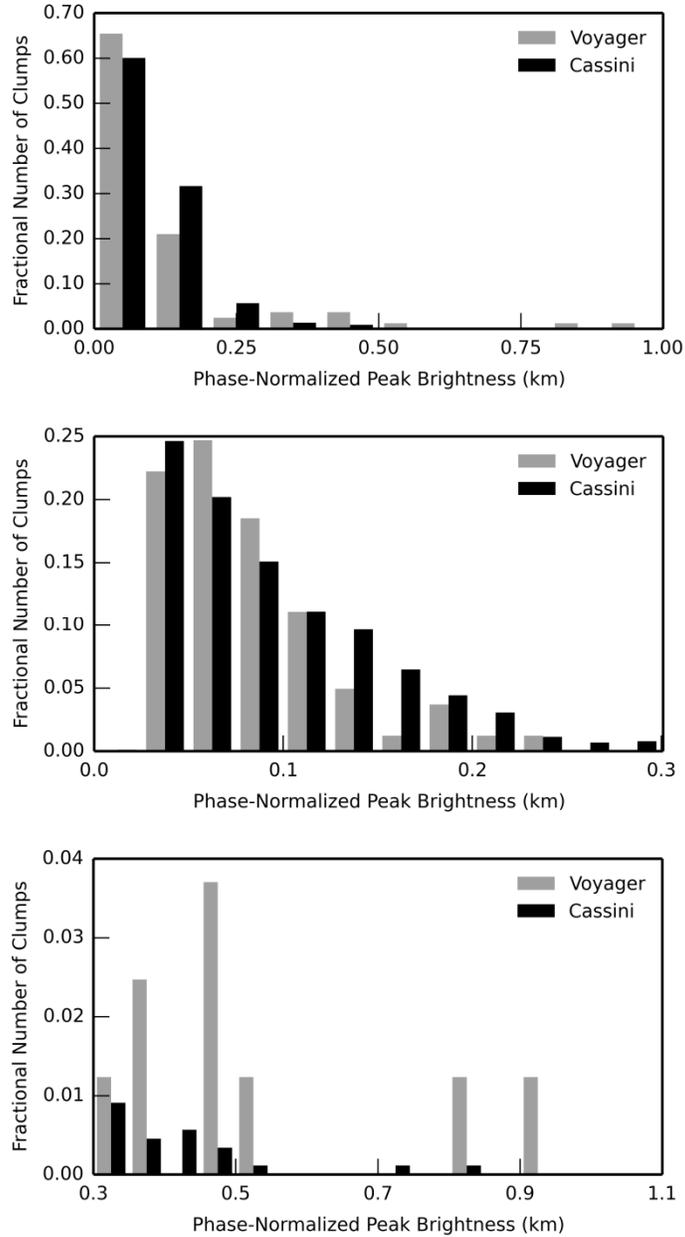

**Figure 12: Distribution of phase-normalized peak brightness for ECs from Voyager and Cassini observations with |$B_0$| > 3°. ECs composing the Cassini C19/2006 and C54/2009 MDCs are not included. The top histogram shows all of the clumps, while the middle and bottom histograms have been zoomed in to show only data on the dimmest and brightest clumps, respectively. The excess of very bright clumps seen by Voyager is apparent.**

The total probability, $V(\lambda)C(\lambda,N)$, of our observations occurring is never greater than ~ 0.01 assuming a constant $\lambda$ (Figure 15). As a result, it is highly improbable that $\lambda$ remained constant between the Voyager and Cassini missions, thus the change in the number of observed bright ECs is significant.

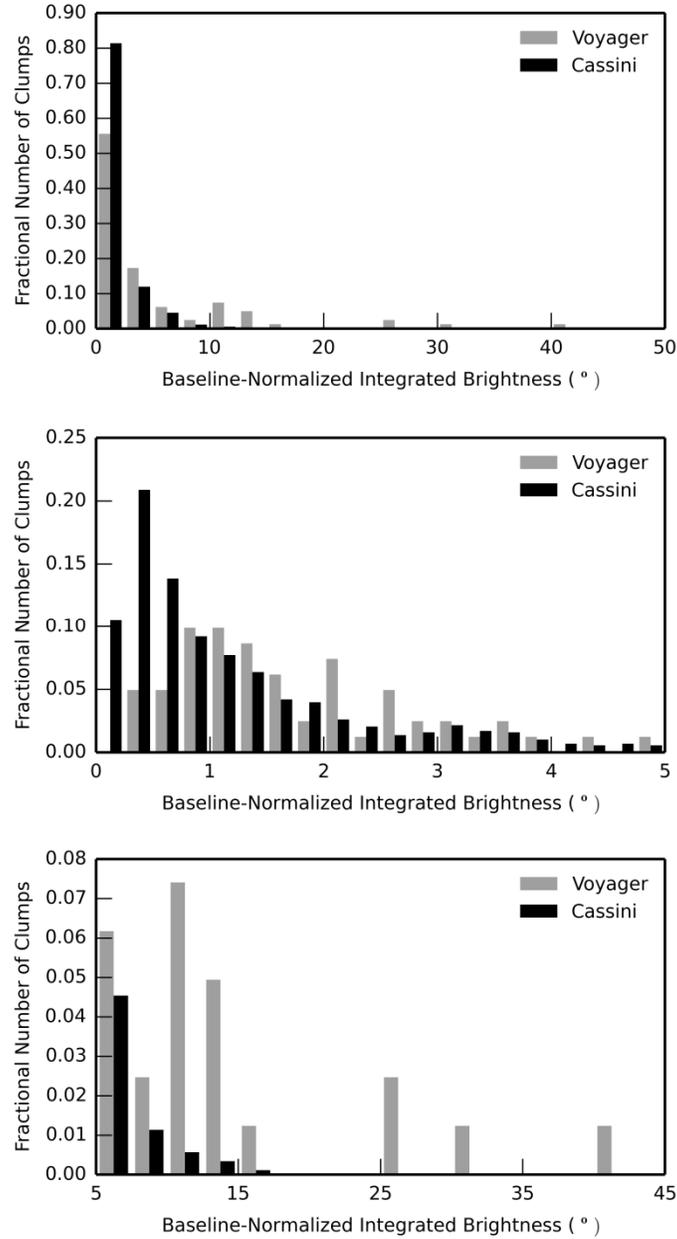

**Figure 13:** Distribution of baseline-normalized integrated brightness for ECs from Voyager and Cassini observations with $|B_0| > 3°$. ECs composing the Cassini C19/2006 and C54/2009 MDCs are not included. The top histogram shows all of the clumps, while the middle and bottom histograms have been zoomed in to show only data on the dimmest and brightest clumps, respectively. The excess of very bright clumps seen by Voyager is apparent.

Finally, we can also examine how brightness relates to clump angular width (Table 3, Figure 16). In all cases, despite significant scatter, the relationship is approximately linear, with increased angular width correlating with increased brightness. Using phase-normalized integrated brightness we find that the best-fit slopes for Voyager and Cassini are approximately equal with small uncertainties; in these cases, a given angular width of clump has approximately the same integrated brightness for both spacecraft. However, for the phase-normalized peak brightness and baseline-normalized cases, the best-fit slopes are considerably different; a clump with a given angular width is two to four times as bright during the Voyager epoch compared to Cassini.

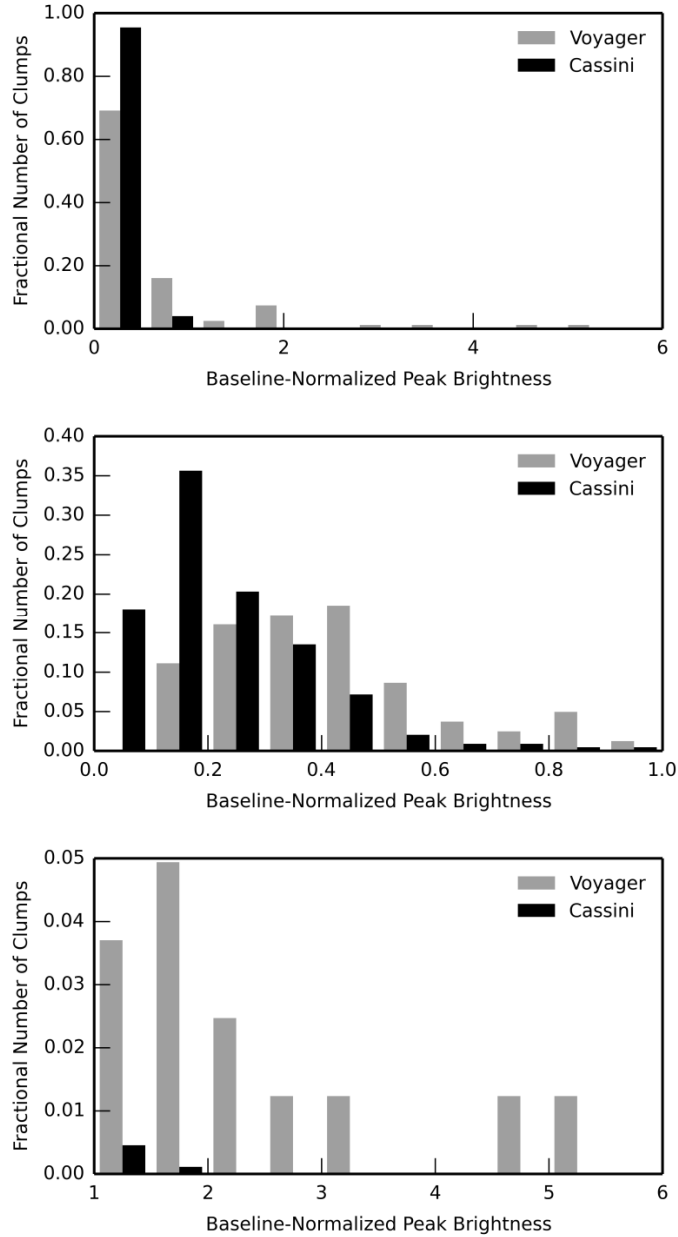

**Figure 14: Distribution of baseline-normalized peak brightness for ECs from Voyager and Cassini observations with $|B_0| > 3°$. ECs composing the Cassini C19/2006 and C54/2009 MDCs are not included. The top histogram shows all of the clumps, while the middle and bottom histograms have been zoomed in to show only data on the dimmest and brightest clumps, respectively. The excess of very bright clumps seen by Voyager is apparent.**

C19/2006, excluded in the above statistics, was dramatically wider in longitudinal extent and brighter than any other Cassini or Voyager clump (Table 3). The difference in baseline-normalized peak brightness is not quite as great, however, as can be seen in Figure 3. The highest peak value (and first observation) for C19/2006 is only slightly higher than the highest peak value for any Voyager clump, and the later peaks for C19/2006 are substantially dimmer than the bright Voyager clumps. The greater difference in baseline-normalized integrated brightness is due to the clump's exceptional angular width and the narrowness of the Voyager clumps.

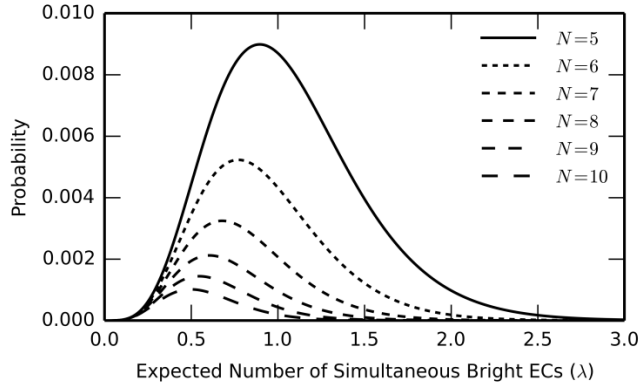

**Figure 15: The probability of each Voyager observing two or more bright ECs and Cassini observing a total of two or fewer bright ECs assuming a constant $\lambda$, the mean number of simultaneous bright ECs. $N$ is the number of distinct Cassini observing periods such that a bright EC is unlikely to survive from one period to the next. Even the most conservative case yields a probability of less than 0.01.**

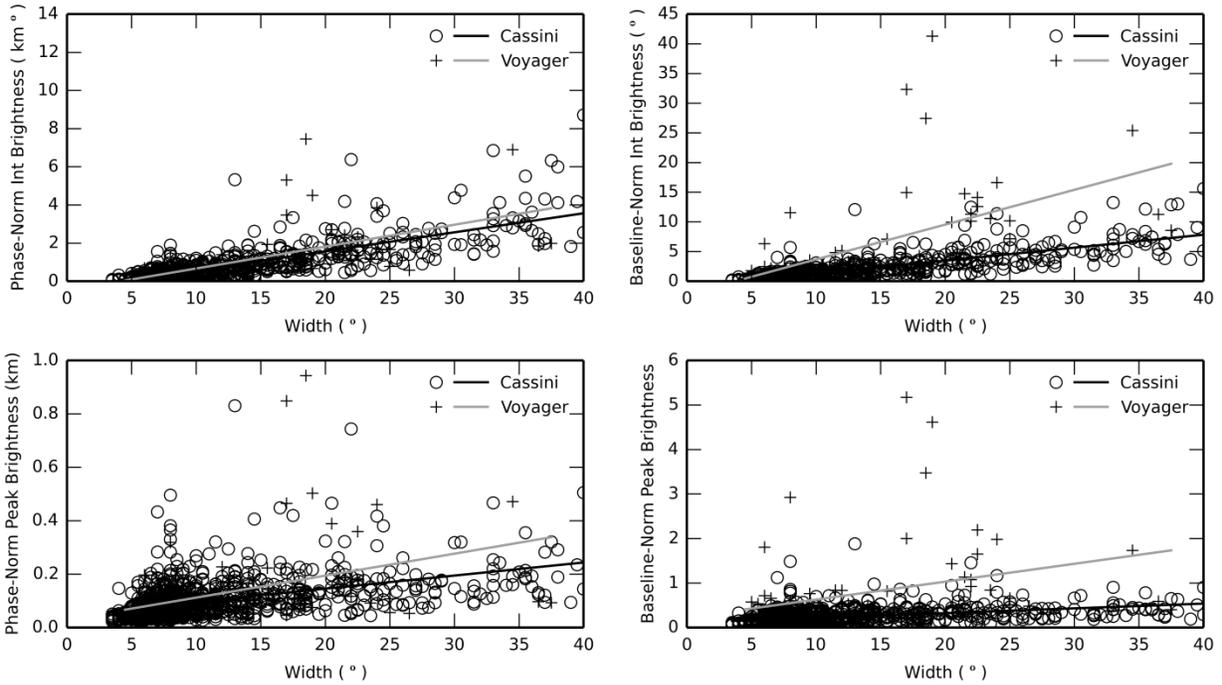

**Figure 16: The relationship between EC angular width and brightness for Voyager and Cassini clumps with $|B_0| > 3°$. ECs composing the Cassini C19/2006 and C54/2009 MDCs are not included. The top plots show integrated brightness and the bottom plots show peak brightness. The best-fit lines are shown and their slopes are given in Table 3.**

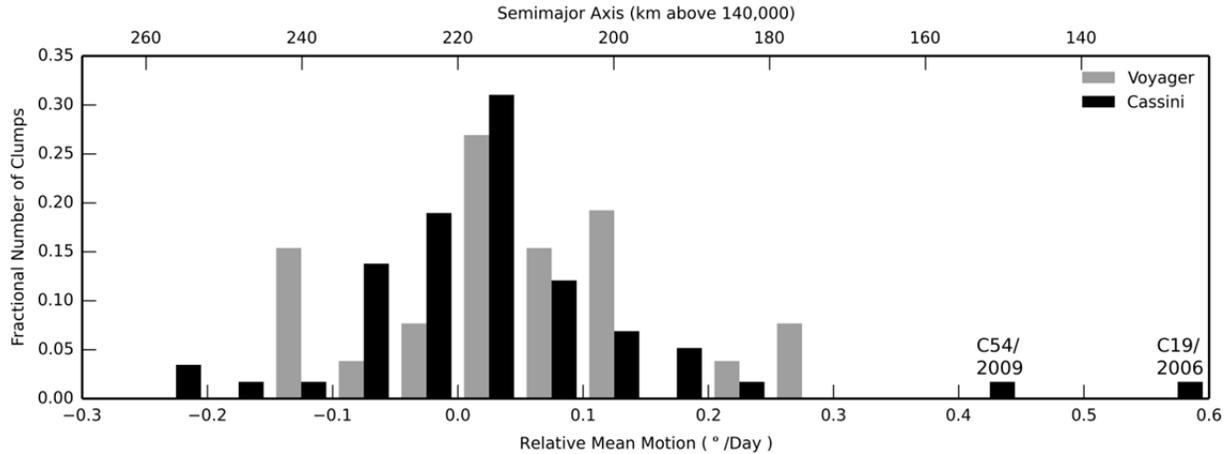

**Figure 17:** Distribution of Voyager and Cassini MDC mean motions relative to the F ring core. The mean motions corresponding to the Cassini C19/2006 and C54/2009 MDCs are marked. The upper axis indicates the semimajor axis corresponding to the given mean motion.

C54/2009, on the other hand, did not show as dramatic a difference (Table 3). While it had somewhat higher integrated brightness than any other Cassini clump, C54/2009 was otherwise similar in physical characteristics. The primary reason this clump was treated differently will be discussed in Section 4.4.

## 4. MULTIPLY-DETECTED CLUMPS

*4.1. Orbit analysis of multiply-detected clumps*

We computed the relative mean motion of each Cassini MDC by finding the slope of the best-fit line to the measured co-rotating longitudes and times. We took the mean motions of the 26 Voyager MDCs directly from S04 as described in Section 2.5. The MDCs we detected in Cassini data, including C19/2006 and C54/2009, had mean motions relative to the F ring core ranging from −0.247 to 0.575 °/day with a mean of 0.024 ± 0.126 °/day. These correspond to a range of semimajor axes of 140,261 to 140,129 km, respectively, ~ 40 km exterior to and ~ 90 km interior to the core. The relative mean motions of the Voyager MDCs from S04 range from −0.134 to 0.296 °/day with a mean of 0.048 ± 0.112 °/day (Table 2). Relative mean motions in both data sets (Figure 17, Table 2) are approximately normally distributed around the core. As we will discuss in Section 4.4, several of the largest measured mean motions may not be due solely to orbital motion, and thus the inferred semimajor axes may be incorrect.

There are many potential sources of uncertainty in our clump longitude measurements that would affect computation of mean motions. Our definition of the center of a clump, which is the center of the Gaussian that best fits the clump profile, is somewhat arbitrary in the case when a clump has a non-Gaussian shape. In these cases, as the clump's shape changes over time, the Gaussian moves relative to the clump and causes changes in measured longitude. In addition, our mosaicked reprojections assume that all objects share a common eccentricity and pericenter. While there is good evidence that the clumps share the eccentricity and pericenter of the F ring core (i.e., clumps are never seen to cross over the core), small differences in eccentricity would lead to changes in observed longitude. Finally, errors in the spacecraft pointing correction used to create the original mosaics could yield incorrect longitudes, but we expect these errors to be on the order of a few pixels at most, resulting in longitude errors less than 0.1°.

To determine the uncertainty for each relative mean motion, when possible we used the standard deviation of the scatter of detected longitudes about a linear best-fit mean motion for each MDC. This is the same technique used by S04, but all of their clumps had many observations, allowing a statistically valid estimate of the uncertainty. The technique works for MDCs with three or more observations, but many of our MDCs only have two observations. Instead, we use MDCs with three or more observations to determine a reasonable uncertainty on our longitude measurements for use when an MDC only has two observations. Looking at 34 MDCs with three or more observations (and ignoring the anomalous C19/2006), we find that the 1-$\sigma$ uncertainty of the observed longitudes about the best fit mean motions is below 0.7° in all cases, with 66% falling below 0.3°. Thus we consider a reasonable assumption of 1-$\sigma$ uncertainty to be 0.3°, and we use this value for all MDCs that have only two observations. Because the uncertainty is relatively small, we do not explore its exact cause from among the possibilities listed above.

The uncertainty in mean motion decreases with increasing length of the time between the first and last observations of a clump. The vast majority of our clumps have baselines of two weeks or more, making the resulting uncertainties quite small – on the order of a few km in semimajor axis.

*4.2. Lifetimes of multiply-detected clumps and clump production rate*

Lifetime measurements are limited by the temporal coverage of our observations: when we first detect a clump, it may have existed since shortly after the previously available observation, and when we last detect a clump it may continue to exist until shortly before the next available observation. Thus the lifetime of each clump must be presented as a range, with the minimum value being the smallest actually observed but the maximum being the largest that might be possible. The occasionally large gaps in time between observations, along with the partial longitude coverage of many observations, mean that the constraints on lifetime may be very weak.

The tightest constraints on lifetime show several clumps that could not have existed for more than ~ 25 days and one clump (not including the long-lived C19/2006) that existed for at least ~ 86 days. Most clumps were actually observed for less than a month. The limited temporal coverage provided by Voyager produced even weaker constraints on clump lifetime. Clumps seen by S04 did not persist during the 9-month time between Voyager missions, but several bright clumps did last for >30 days with only minor changes. S04 interpreted this to mean that the clumps may continue to live much longer. McGhee et al. (2001) and Bosh and Rivkin (1996) found similar results during the 1995 ring plane crossing, with observed clump lifetimes up to ~ 3 months (and less than ~ 6 months) and typical lifetimes of ~ 1 month. Unfortunately, we have insufficient observations to determine whether clump lifetimes have changed between 1980 and 2010.

The number of ECs present at any one time, combined with an approximation of clump lifetime, allows us to compute the rate of clump production. While counting ECs, we use only observations with more than 80% longitudinal coverage. For those observations with more than 80% but less than 100% longitudinal coverage, we weight the number of clumps by the fractional coverage. C19/2006 and C54/2009 are included, as are observations near the equinox.

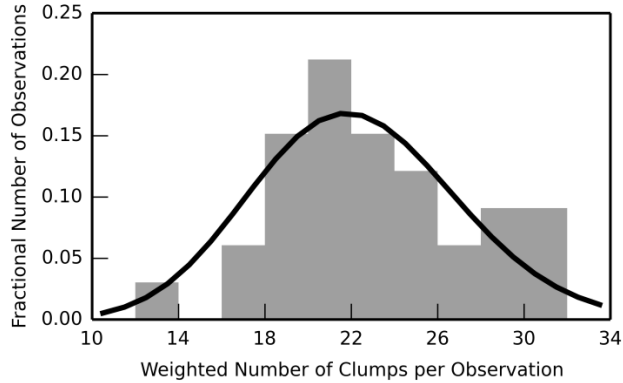

**Figure 18: Distribution of number of ECs per observation for Cassini. ECs composing the Cassini C19/2006 and C54/2009 MDCs are included. The best-fit Poisson distribution (mean 22.3) is shown.**

Cassini saw between 13 and 33 clumps at any given time, with a mean of $23.2 \pm 4.7$. Voyager saw between 20 and 22 clumps, with a mean of $20.9 \pm 1.1$, but with a much smaller temporal coverage and only four observations. The mean numbers of clumps are consistent within $1\sigma$. From this we conclude that the number of clumps at any given time during the Cassini mission is approximately the same as it was during the Voyager missions. However, while Voyager saw 2–3 bright, large clumps at any given time (in all of our limited snapshots), Cassini only sees one such clump (like the C19/2006 and C54/2009) at a time, and then only once every few years.

The distribution of the number of Cassini clumps is shown in Figure 18. The distribution is roughly Poisson with a mean of 22.3. If we assume the average clump lifetime is ~ 45 days, then clumps are produced (and destroyed) at a rate of ~ 0.5 per day. The Poisson-like shape of the distribution supports the notion that F ring clumps are independent, statistically uncorrelated events.

*4.3. Appearance, evolution, and disappearance of multiply-detected clumps*

Due to the difficulty of identifying any physical process affecting a large longitudinal segment of the ring simultaneously, it is reasonable to assume that clumps are formed with relatively narrow angular widths but non-zero radial extent, perhaps through local gravitational aggregation or the collision of a small moonlet with the core, and then grow through Kepler shear. A clump with radial extent $\Delta a$ will grow in longitude by an amount $\Delta w = 3/2(n/a)\Delta a \Delta t$, where $n = 581.979$ °/day is the mean motion of the F ring core and $a = 140,220$ km is the semimajor axis of the core. Thus for every km in radial extent, a clump will spread by ~ 0.0062 °/day.

We examined the change in angular width over time for 42 MDCs for which we had at least two weeks' worth of observations (Figure 19). We limited ourselves to these longer time baselines because the dependence of angular width on clump shape discussed in Section 2.4 is magnified when small changes in shape are divided by only a few days. For each MDC we fit a linear slope to the available angular width measurements. Our measured rates of angular width change generally range from –0.20 °/day to 0.19 °/day with three outliers (C19/2006, C22, and C54/2009) above 0.6 °/day that will be discussed below. The growth rate of 0.19 °/day corresponds to a radial extent of ~ 30 km, or about six pixels in our mosaics. Such a clump, surviving for a month, would only grow ~ 6° during its lifetime.

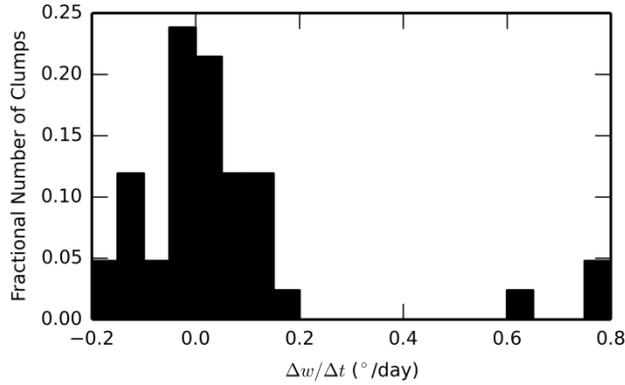

**Figure 19: Change in angular width per day for the 42 multiply-detected Cassini clumps with more than 2 weeks of observations.**

The rates of angular width change (including the outliers) have a mean of 0.05 ± 0.21 °/day and a median of 0.01 °/day. The small growth rate is consistent with the observation by S04 that bright clumps tended to remain approximately the same size over their observed lifetimes. However, the median of 0.01 °/day and the negative growth rates show, in addition, that clumps are about as likely to *shrink* as they are to grow. Visual examination of the shrinking clumps reveals two primary causes. The first is the dependence of our angular width measurement on clump shape (Section 2.4). A clump can change shape in subtle ways, causing our Gaussian to fit with a smaller width. The second cause is more interesting – interaction with Prometheus. In the majority of shrinkage cases, there is evidence of a passage of Prometheus during the clump's lifetime. The 3.2° period of the channel-streamers caused by this passage tends to "chop up" clumps and reduce their angular width (Figure 20).

The generally slow rate of angular width change makes our suggested formation mechanism problematic. If a clump can only grow by a few degrees over its lifetime, how do we regularly see clumps that are tens of degrees in extent? Luckily, the long baseline of our observations allows us to occasionally observe the ring shortly before a clump appears or shortly after a clump disappears. As an example, the appearance of clump C33 is shown in Figure 21. Two observations of this clump, approximately one week apart, are available. C33 grows in angular width from 4.5° to 4.8° in 7.1 days, a change of only ~ 0.05 °/day. If we assume Kepler shear as discussed above, C33 would have had an angular width of ~ 3.3° in the previously available profile ~ 3 weeks earlier. While this size is slightly below our wavelet detection threshold, the clump would nevertheless have been obvious in the previous profile and mosaic (Figure 21, top panel), and no such clump is seen. Unfortunately, the next available observation is nearly four months later, preventing us from seeing how this clump eventually disappears.

Similarly, Figure 22 shows the disappearance of clump C12. This clump grows from 7.1° to 7.8° in 12.0 days, a change of ~ 0.06 °/day. C12 would have had an angular width of ~ 8.6° in the next available profile 13 days later, but no such clump is seen.

- 23 -

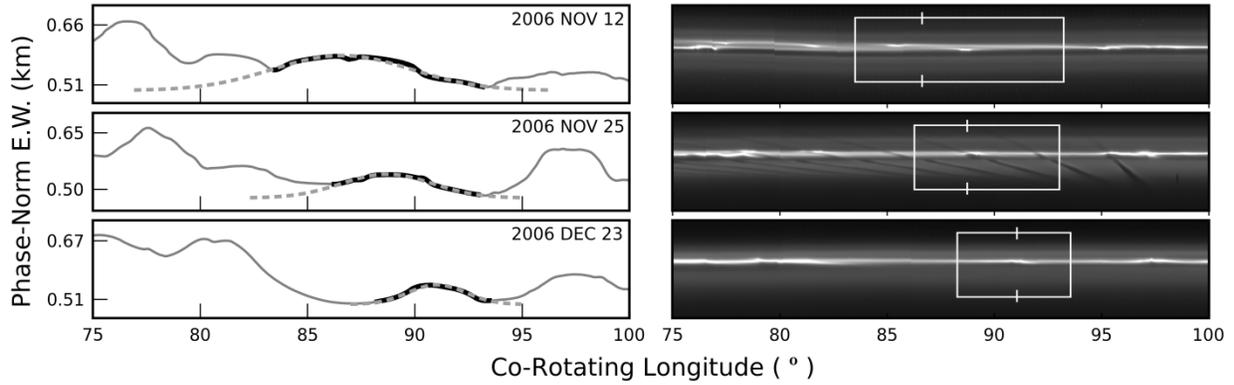

**Figure 20: The evolution of clump C17, with profiles (left) and corresponding mosaics (right). The mosaics have been contrast enhanced for better visibility. The white box indicates the lateral extent of the clump and the white tick marks indicate the center of the Gaussian fit. The passage of Prometheus is indicated by the streamer-channels present in the second panel, and C17 narrows during and after its passage. The angular widths from top to bottom are 9.7°, 6.8°, and 5.3° yielding a net growth rate of –0.11 °/day.**

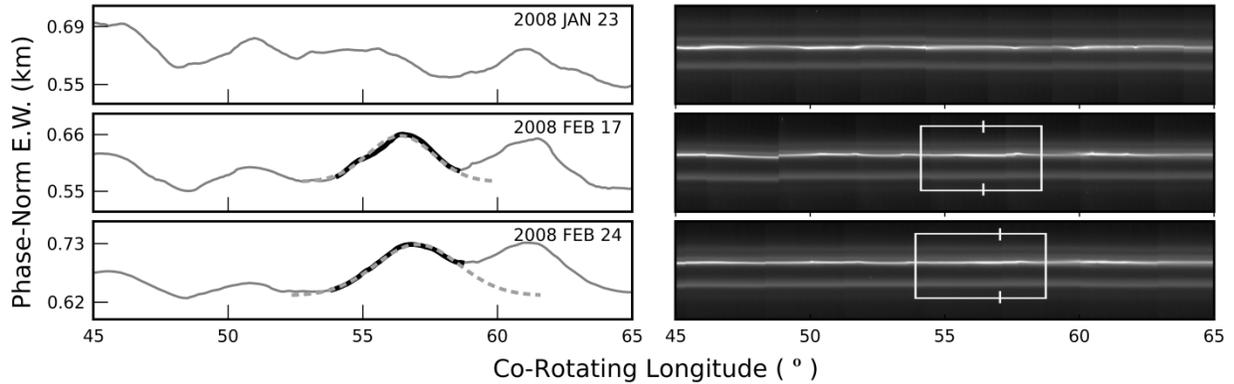

**Figure 21: The appearance of clump C33, with profiles (left) and corresponding mosaics (right). The mosaics have been contrast enhanced for better visibility. The white box indicates the lateral extent of the clump and the white tick marks indicate the center of the Gaussian fit. The top panel is the previously available observation before the clump was first observed.**

Changes of ~ 0.05 °/day correspond to a radial extent of only ~ 8 km, less than two pixels at our mosaic resolution. Thus it is not surprising that we do not observe a substantial radial thickening of the ring in these cases. However, what is surprising is that the clumps seem to appear from the background already near their full size, and later disappear into the background without significant spreading occurring in between.

*4.4. Unusual clumps*

There are three clumps in Table 2 that show exceptionally large angular width growth rates. C19/2006 has a growth rate of 0.63 °/day. C54/2009 has a growth rate of 0.77 °/day. C22 has not been previously identified as an anomalous feature but has a growth rate of 0.79 °/day. Assuming Kepler shear, these growth rates would correspond to spreads in semimajor axes of ~ 100 km. C19/2006 and C54/2009 are also unique in that they have two of the largest mean motions relative to the core: 0.575 °/day for C19/2006 and 0.418 °/day for C54/2009. If these large relative mean motions are due solely the clump orbits, the clumps would be 80–100 km interior to the core.

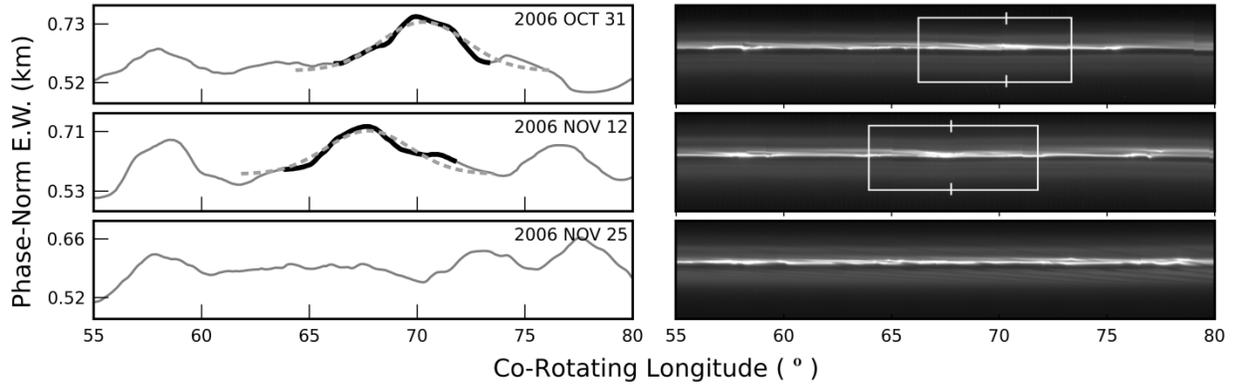

Figure 22: The disappearance of clump C12, with profiles (left) and corresponding mosaics (right). The mosaics have been contrast enhanced for better visibility. The white box indicates the lateral extent of the clump and the white tick marks indicate the center of the Gaussian fit. The bottom panel is the next available observation after the clump was last observed.

While clearly some part of clump growth is due to Kepler shear, the observed radial extents of C19/2006 and C54/2009 are insufficient to explain their growth rates fully. At the same time, we do not find these clumps sufficiently interior to the core to explain their high mean motions. Murray et al. (2008) investigated C19/2006 and determined that, in each available observation, the location of the "disturbance" that we identify as the center of the clump corresponds closely to the position of the small moonlet S/2004 S6. S6 has a semimajor axis $a$ = 140,134.581 km and an eccentricity $e$ = 0.00183319, sufficient to cause it to cross the F ring core on each orbit. S6's orbit has a mean motion of 582.521196 °/day, which is ~ 0.542 °/day relative to the core, similar to the relative mean motion we find for C19/2006. Thus it appears that S6 collided with the core material on multiple orbits, causing a progression of dust clouds that we interpret as a clump spreading quickly and moving rapidly with respect to the core.

C54/2009 appears to be similar, in that its spreading rate and mean motion can not be fully explained by its location and radial extent. In some of the mosaics a "mini-jet" with a "bright head" (Attree et al., 2014) is clearly visible. The bright head may be the body that is colliding with the core over multiple orbits, but we do not have enough observations to derive an orbit for this moonlet.

In contrast to the above, our third rapidly spreading clump (C22, Figure 23) has a mean motion of only –0.010 °/day. It, too, has an obvious mini-jet with a bright head, indicating that it is likely formed by a collision as well. However, in this case, there appears to be only a single collision, after which the moonlet was either destroyed or moved to a different orbit. C22 does have significant radial extent, which is able to explain its rapid spreading. A spread rate of 0.79 °/day corresponds to a radial extent of ~ 125 km, or about 13 pixels on either side of the core, which is roughly what is seen in the mosaics.

The existence of these three unusual clumps suggests that there may be two distinct populations of clumps. The most common clumps appear at full size, change little, and then fade away. The second, more rare type is formed by the (possibly repeated) collision of a small moonlet with the F ring core. These clumps have substantial radial extent and may appear to move with the location of the moonlet's crossing of the core.

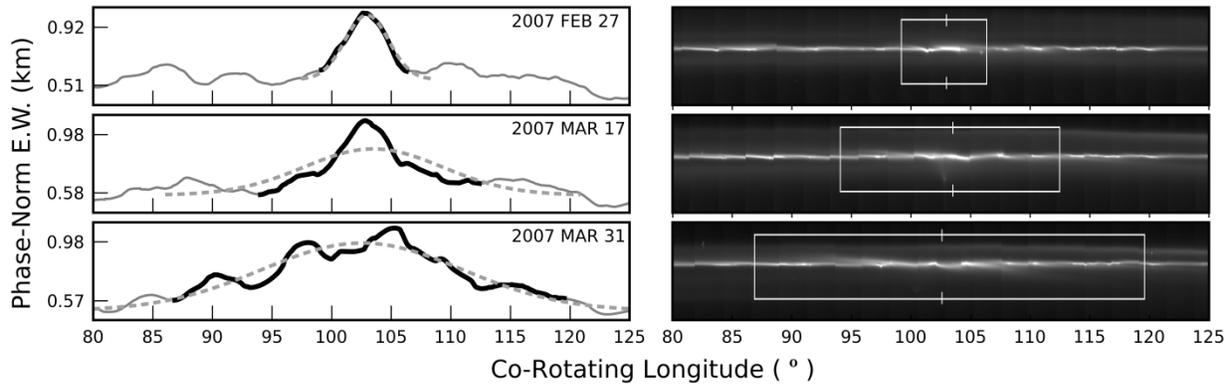

Figure 23: The evolution of clump C22, with profiles (left) and corresponding mosaics (right). The mosaics have been contrast enhanced for better visibility. The white box indicates the lateral extent of the clump and the white tick marks indicate the center of the Gaussian fit. The angular widths from top to bottom are 7.2°, 18.4°, and 32.7° yielding a net growth rate of 0.79 °/day.

The unusual clumps seen by the Voyager missions appear to be fundamentally different. As noted by S04 and clearly visible in Figure 3, Voyager saw 2–3 exceptionally bright clumps at any given time. The clumps seen by Voyager 1 lasted, essentially unchanged, for more than a month but did not survive the nine months between missions, indicating that new bright clumps must have formed in the interim. Thus, the rate of production of these bright clumps must be on the order of several per year. However, we do not see bright, unchanging clumps in the Cassini data (Figure 3). All bright clumps either move rapidly with respect to the core (indicating successive collisions with a moonlet) or spread rapidly (indicating a large radial extent). It appears that the bright Voyager clumps may have been the result of single collisions at low speeds, such that a large amount of dust was released but not with large radial spreads.

Showalter (1998) did detect three clumps that appeared rapidly, spread at ~ 0.3 °/day, and disappeared within two weeks. He attributed these clumps to hypervelocity impacts of centimeter-sized meteoroids that caused a ~ 100 km radial spread of ring material. Barbara and Esposito (2002) disagreed with this interpretation, primarily on the basis that the many smaller events that they expected to occur were not seen. They instead proposed that the burst events were caused by disruptive collisions involving poorly consolidated moonlets. This latter explanation seems more likely for the three bright clumps seen by Cassini, particularly now that we know that objects like S/2004 S6 exist. However, the events seen by Cassini were also dramatically brighter and much longer-lived.

Finally, S04 noted that their clump 2C′ was associated with a series of clumps that appeared to diverge from its location at different points in time. This is consistent with the interpretation of a moonlet colliding with the core at different locations, spawning multiple clumps, each with its own mean motion.

In summary, unusual bright clumps can expand in angular width due to their radial extent, expand in angular width due to repeated collisions with a moonlet that has a semimajor axis different from that of the core, or not expand in angular width. The first two cases have been seen in the Cassini data, but are rare, with an occurrence of only one every few years. All three cases have been seen in the Voyager data, but were very common, with an occurrence of at least several per year. This difference suggests a fundamental change in the population of moonlets in the vicinity of the F ring.

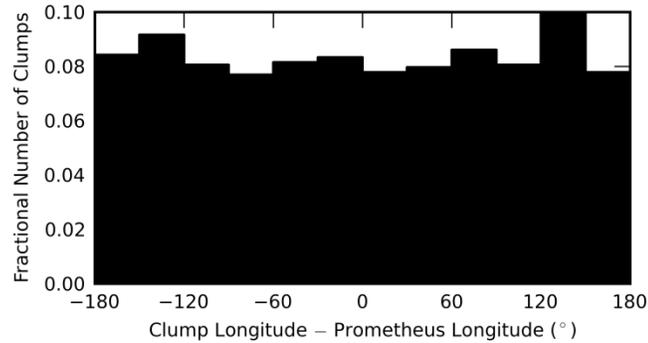

**Figure 24: Distribution of difference between clump location and the location of Prometheus for all Cassini ECs. No correlation is seen.**

## 5. POSSIBLE INFLUENCES ON CLUMP FORMATION

*5.1. The role of Prometheus*

  The F ring is bracketed by the orbits of two "shepherd" moons, Prometheus and Pandora. Both Prometheus and the F ring are on eccentric orbits, resulting in a closest approach every orbit (~ 14.7 hours). The relative mean motions of Prometheus and the F ring result in the point of closest approach moving ~ 3.2° along the core each orbit, leaving characteristic features such as streamer-channels in the F ring material (Beurle et al., 2010; Chavez, 2009; Giuliatti Winter et al., 2000; Kolvoord et al., 1990; Murray et al., 2008, 2005; Showalter and Burns, 1982). As we are interested in the much larger clumps, we explicitly filter these features out as described in Section 2.4. In contrast, only a hint of the ~ 5.75° pattern induced by Pandora has been found in Cassini images, with no evidence found in the Voyager images, and we do not enlarge our filter to remove these possible effects (Kolvoord et al., 1990; Porco et al., 2005). The influence of Prometheus on the F ring is so great that it is tempting to attribute clump formation to its presence.
  To determine what influence Prometheus may have, we have compared the position of each EC with the position of Prometheus at that time. Prometheus's 67.6-day synodic period is comfortably longer than our estimated lifetimes, implying that Prometheus will not generally "lap" a clump. If Prometheus influences clump creation, we would expect to see a clustering of clump longitudes behind Prometheus' position and fewer in front.
  We look for such a clustering by measuring the distance in longitude between each EC and Prometheus' position at the time it was observed (Figure 24). We find no dependence of clump location on Prometheus, indicating that Prometheus plays no direct role in clump formation.

- 27 -

*5.2. The role of embedded moonlets*

It is well-accepted that the region surrounding the F ring is home to a large number of small bodies (Albers et al., 2012; Beurle et al., 2010; Cuzzi and Burns, 1988; Meinke et al., 2012; Murray et al., 2008; Porco et al., 2005). These "moonlets" may be located in the F ring core or may be on orbits slightly different from the core, causing them to pass through the core each orbit. The largest of these moonlets, such as S/2004 S6, may collide with the core multiple times, while smaller moonlets may only collide with the core once or a handful of times before being destroyed. Loose aggregations of material and moonlets can collide in high-speed collisions (~ 30 m/s) that result in highly visible dust plumes or "jets". These jets can eventually produce kinematic spirals due to Kepler shear (Charnoz, 2009; Charnoz et al., 2005). Low-speed collisions (~ 1 m/s) may only result in localized dust clouds or small features such as "mini-jets" (Attree et al., 2014).

In Section 4.4, we discussed the role of high-speed impacts on the formation of unusual clumps. We also investigated whether the clumps are associated with the smaller features caused by low-speed impacts. Attree et al. (2014) have cataloged 889 of these small ($\ll 1°$ in longitude) F ring features observed by Cassini. Of these, 437 are "mini-jets", short-lived (~1 d) and small (~ 50 km) linear features protruding from the F ring core that appear to be the result of low-speed impacts of < 1 km radius objects; 245 of these features are "complex", difficult-to-analyze structures that may contain multiple mini-jets and other structure. The remaining 207 are "extended objects", bright features close to, but not coincident with, the F ring core. At any given time there are ~ 15 examples of these three types of features visible in the ring and they are distributed randomly with no apparent large-scale clustering.

Forty-five of our 65 observations were also analyzed by Attree et al. (2014) and were found to have at least one feature. Attree et al. (2014) did not claim to identify every such feature in each observation, only a representative set sufficient to perform statistical analysis. As such, to correlate these features with our ECs, we first look at each feature and then determine whether it is inside a clump. Of the 889 total features, we have profile data for 516. Of these, 451 (87%) are associated with an EC. Limiting ourselves to only mini-jets does not change this percentage. Overall, ECs cover 54–95% of each profile, with a mean of 78 ± 9%. Thus, if the features are randomly distributed and are not correlated with the ECs, we would expect ~ 78% of the features to be associated with a clump. The actual association is one standard deviation above this expected value, suggesting that in fact these features may be more likely to appear in our ECs. However, this evidence is inconclusive.

## 6. DISCUSSION

*6.1. Clump Origins*

To summarize our major results, Voyager and Cassini saw approximately the same number of clumps at any given time. The angular widths, absolute brightnesses, lifetimes, and semimajor axis distributions of clumps are roughly similar within the limits of our observations. Clumps have radial spreads of 0–100 km and may change size due to Kepler shear or due to the effects of the passage of Prometheus, although many clumps remain the same size during their observed lifetimes. Most clumps form already near their mature size and then fade back into the background ring core material over time, but a few, including the two brightest clumps in the Cassini data, show distinct spreading due to Kepler shear. The location of clumps does not appear to have any relation to the position of Prometheus. There is only a loose association between the extended clumps we have investigated here and the small-scale features such as mini-jets discussed by Attree et al. (2014) and others.

Because some clumps spread rapidly, apparently due to Kepler shear, but others do not, we hypothesize that ECs are the consequence of diverse physical processes. Charnoz et al. (2009) proposed that the 2–3 large, bright clumps seen by Voyager at any given time were the result of high-speed collisions by embedded moonlets. These collisions are capable of spreading ring material over several hundred km in semimajor axis. It seems likely that such clumps would be similar to C19/2006 and C54/2009 seen by Cassini, and the remnants of another large impact were apparent in the kinematic spiral observed early in the Cassini tour (Charnoz et al., 2005). Showalter (1998) did note two Voyager-era clumps that appeared suddenly and sheared rapidly. However, the brightest Voyager clumps did not shear rapidly, so Charnoz et al.'s inference is unverified.

What of the more stable clumps? S04 suggested that a typical clump could be formed from the total destruction of a body ~ 12 m in radius or the release of regolith 10 cm deep on a body ~ 80 m in radius. In other words, a bright clump need not comprise a large volume of ring material. However, the question remains how such an impact could result in a longitudinally extended bright region, when the radial extent is too small for Kepler shear to play a significant role. S/2004 S6 has been seen in some images to consist of an extended chain of small objects, which when impacting the core could produce a similarly extended bright region. However, even if true for S6, this mechanism is unlikely to be common.

Beurle et al. (2010) have suggested that Prometheus plays a double role when influencing the F ring. During its closest approach to the F ring core each orbital period, Prometheus perturbs a small (~ 1°) segment of the ring, forming streamer-channels, and repeats the process 3.27° further along on the next orbit. The perturbations induce the formation of small, dense objects at channel edges. Later passes of Prometheus can disrupt or enhance these objects, eventually resulting in the formation of moonlets or small clumps. Some of these objects have also been shown to survive future passages of Prometheus intact.

We propose that the stable clumps are likewise a delayed effect of the passage of Prometheus. The immediate perturbations caused by the latest passage of Prometheus shear out, and clumps arise from the secondary collisions that ensue. These resulting clumps are thus extended in longitude when they appear, but are disassociated from the current location of Prometheus. This is similar to the delayed clump formation predicted by Showalter and Burns (1982).

*6.2. Changes 1980–2010*

As noted above, many of the properties of F ring clumps are similar in the Voyager and Cassini data sets. However, the ring *looks* very different (Fig. 3). There are two reasons why. First, the F ring's baseline was much lower in the Voyager era (French et al., 2012), making smaller clumps much more visible to the eye. The other is that, although the populations of the numerous smaller clumps are similar, the F ring has shown a marked decrease in its population of the very brightest clumps. Only one clump observed by Cassini has ever exceeded the brightness of the brightest Voyager-era clumps, and it did so for only a few months in 2006–2007.

The change in the properties of the largest ring clumps needs to be explained. Specifically, large impacts are infrequent but energetic in the Cassini data, whereas they were more common but much more gentle in the Voyager data. French et al. (2012) cataloged several other major changes in the F ring from the Voyager era to the present: the ring is both brighter and wider today, suggesting that it is dynamically hotter.

Consider the following scenario. Suppose that, in the Voyager era, the F ring was surrounded by a population of moonlets or clumps on similar orbits. The existence of this population is supported by charged particle absorption signatures near the ring in Pioneer and Voyager data (Cuzzi and Burns, 1988). By "similar orbits", we assume that pericenters are nearly aligned, so that orbits intersect slowly or not at all; this is supported by the study of the F ring's strands in the Voyager data by Murray et al. (1997).

Over a time scale of decades, several fundamental changes would be expected to take place. First, the number of moonlets would be depleted as they collide with the ring's core. Second, the ring would heat up due to all of these collisions. Third, a time scale of decades is sufficient for orbital pericenters to process out of alignment, so that later impacts would occur at higher speeds. Fourth, a time scale of decades might also be sufficient for members of the impactor population accrete into larger and more visible moonlets like S/2004 S6. Such a scenario is reminiscent of the predator-prey model of Esposito et al. (2012) and can account for most of the key observed properties of F ring clumps during the last 25–30 years.

However, what process could have generated the moonlet population on a time scale of a few decades? Although a single, unusually large impact might have created the neighboring swarm shortly before 1980, this seems statistically unlikely (but not impossible). An alternative explanation is that the process is cyclic, related to the 17-year cycle defined by the mutual precession rate of Prometheus relative to the ring (Borderies and Goldreich, 1983). Prometheus' influence on the ring is strongest when its pericenter is anti-aligned with that of the F ring, so that Prometheus enters deeply into the ring at each pericenter passage. Anti-alignments occurred in 1975, 1992 and 2009.

Prometheus' effect on the ring is strong but localized; it does not stir up the ring uniformly. Instead, at each pericenter it casts a small number of ring particles out into the surrounding region. The Voyager encounters occurred a few years after the 1975 anti-alignment, when impacts by nearby objects were numerous but slow. Our subset of the Cassini data set spans 12–17 years after the 1992 anti-alignment, when the ring is hotter and collisions are faster but less frequent. Dynamical modeling will be required to test this hypothesis in detail, but it could plausibly account for all of the observations. It implies that the F ring's changes are seasonal but with a time lag, much as the seasonal temperature variations on Earth lag behind the Sun's elevation in the sky.

Winter et al. (2007) proposed an additional mechanism that could contribute to the impactor population. They found that test particles in the F ring wandered on chaotic orbits due to the influence of Prometheus and Pandora, which themselves are on chaotic orbits (Cooper and Murray, 2004; Farmer and Goldreich, 2006; French et al., 2003; Goldreich and Rappaport, 2003a, 2003b; Poulet and Sicardy, 2001). These particles experienced jumps in semimajor axis of up to ~ 100 km, placing them on noticeably different orbits than the F ring core.

If the moonlet population is seasonal, then we would expect to see a return to Voyager-like clump formation in the next few years. If Prometheus is not involved, then the rate of clump formation should continue to diminish. Future observations will distinguish these cases for sure.

In this paper, by viewing clumps at the multiple-degree scale and eliminating any radial information, we have lost some clues to the characteristics and origins of particular clumps. Clumps may, in fact, be quite different in origin and structure, but still look similar when viewed as an equivalent width profile. This work therefore serves as a complement to the more detailed analyses by Murray et al. (2008), Beurle et al. (2010), and Attree et al. (2014).

Unfortunately, our approach appears to be the only practical way to study the long-term changes in the F ring. The limited temporal spacing of Cassini's F ring observations prevents us from seeing the origin and evolution of most clumps, whereas Voyager's limited spatial resolution prevents us from seeing the clumps' internal structure. Equivalent width profiles and associated derived data therefore provide our best means for studying the changes between the two eras.

## ACKNOWLEDGMENTS

This work was funded by NASA's Cassini Data Analysis Program through Grants NNX07AJ76G and NNX09AE74G and the NSF REU program Grant #AST0852095.

**Table 1: Summary of Cassini images used.** All images are part of an image sequence with first and last image numbers as shown. ISS_085RF_FMOVIE003_PRIME consists of images taken at one ansa for a period of time followed by the other ansa. We split this observation to isolate each view. Emission and incidence angles are relative to the normal vector on the lit side of the ring plane. Coverage is the percentage of 360° covered by the available data. # of MDCs is the number of MDCs that have a detection in the given observation.

| Obs # | Date | Cassini Observation ID | Radial Resolution (km/pixel) | Phase Angle (°) | Emission Angle (°) | Incidence Angle (°) |
|---|---|---|---|---|---|---|
| 1 | 2004 JUN 20 | ISS_000RI_SATSRCHAP001_PRIME | 38 – 86 | 67.1 – 67.2 | 73.9 – 73.9 | 65.5 |
| 2 | 2004 NOV 15 | ISS_00ARI_SPKMOVPER001_PRIME | 27 – 43 | 84.3 – 84.9 | 77.2 – 77.3 | 66.6 |
| 3 | 2005 APR 13 | ISS_006RI_LPHRLFMOV001_PRIME | 6 – 17 | 31.2 – 39.4 | 83.2 – 83.7 | 68.0 |
| 4 | 2005 MAY 01 | ISS_007RI_LPHRLFMOV001_PRIME | 6 – 8 | 28.5 – 35.6 | 68.4 – 69.4 | 68.2 |
| 5 | 2005 MAY 02 | ISS_007RI_AZSCNLOPH001_PRIME | 29 – 66 | 5.5 – 25.6 | 62.7 – 69.4 | 68.2 |
| 6 | 2005 MAY 03 | ISS_007RI_HPMRDFMOV001_PRIME | 4 – 13 | 117.2 – 130.1 | 93.8 – 97.7 | 68.2 |
| 7 | 2006 SEP 28 | ISS_029RF_FMOVIE001_VIMS | 9 – 10 | 158.4 – 161.8 | 120.9 – 122.5 | 74.3 |
| 8 | 2006 SEP 30 | ISS_029RF_FMOVIE002_VIMS | 11 – 12 | 158.9 – 161.0 | 118.5 – 119.4 | 74.3 |
| 9 | 2006 OCT 16 | ISS_030RF_FMOVIE001_VIMS | 10 – 11 | 150.6 – 152.4 | 130.3 – 131.3 | 74.6 |
| 10 | 2006 OCT 31 | ISS_031RF_FMOVIE001_VIMS | 9 – 10 | 156.3 – 160.3 | 124.8 – 128.1 | 74.8 |
| 11 | 2006 NOV 12 | ISS_032RF_FMOVIE001_VIMS | 9 – 10 | 155.9 – 160.0 | 124.9 – 128.2 | 74.9 |
| 12 | 2006 NOV 25 | ISS_033RF_FMOVIE001_VIMS | 10 – 10 | 159.8 – 160.7 | 119.3 – 122.4 | 75.1 |
| 13 | 2006 DEC 23 | ISS_036RF_FMOVIE001_VIMS | 12 – 12 | 158.7 – 160.8 | 122.8 – 125.7 | 75.5 |
| 14 | 2007 JAN 05 | ISS_036RF_FMOVIE002_VIMS | 10 – 11 | 131.2 – 135.6 | 143.8 – 145.5 | 75.7 |
| 15 | 2007 FEB 10 | ISS_039RF_FMOVIE002_VIMS | 10 – 10 | 125.5 – 131.1 | 146.9 – 148.6 | 76.2 |
| 16 | 2007 FEB 27 | ISS_039RF_FMOVIE001_VIMS | 10 – 10 | 101.6 – 108.1 | 143.8 – 146.6 | 76.4 |
| 17 | 2007 MAR 17 | ISS_041RF_FMOVIE002_VIMS | 10 – 11 | 106.2 – 111.8 | 143.6 – 145.3 | 76.7 |
| 18 | 2007 MAR 31 | ISS_041RF_FMOVIE001_VIMS | 12 – 12 | 82.2 – 85.6 | 128.2 – 130.3 | 76.9 |
| 19 | 2007 APR 18 | ISS_043RF_FMOVIE001_VIMS | 12 – 12 | 83.8 – 87.3 | 125.6 – 127.5 | 77.1 |
| 20 | 2007 MAY 05 | ISS_044RF_FMOVIE001_VIMS | 12 – 12 | 80.2 – 83.4 | 118.6 – 120.0 | 77.4 |
| 21 | 2007 OCT 18 | ISS_051RI_LPMRDFMOV001_PRIME | 32 – 85 | 55.0 – 56.9 | 94.0 – 94.1 | 79.8 |
| 22 | 2007 DEC 31 | ISS_055RF_FMOVIE001_VIMS | 9 – 10 | 62.8 – 67.7 | 122.1 – 123.6 | 81.0 |
| 23 | 2008 JAN 07 | ISS_055RI_LPMRDFMOV001_PRIME | 11 – 14 | 19.2 – 22.6 | 99.7 – 102.6 | 81.1 |
| 24 | 2008 JAN 23 | ISS_057RF_FMOVIE001_VIMS | 10 – 10 | 42.5 – 46.1 | 117.2 – 119.4 | 81.3 |
| 25 | 2008 FEB 17 | ISS_059RF_FMOVIE001_VIMS | 8 – 9 | 43.1 – 46.9 | 121.5 – 124.2 | 81.7 |
| 26 | 2008 FEB 24 | ISS_059RF_FMOVIE002_VIMS | 9 – 10 | 24.1 – 27.1 | 104.9 – 108.7 | 81.8 |
| 27 | 2008 MAR 15 | ISS_061RI_LPMRDFMOV001_PRIME | 8 – 25 | 12.6 – 15.1 | 94.2 – 97.1 | 82.1 |
| 28 | 2008 JUN 14 | ISS_072RI_SPKHRLPDF001_PRIME | 81 – 111 | 32.5 – 39.4 | 115.8 – 122.8 | 83.5 |
| 29 | 2008 JUL 05 | ISS_075RF_FMOVIE002_VIMS | 6 – 7 | 28.9 – 37.1 | 112.4 – 120.9 | 83.8 |
| 30 | 2008 JUL 08 | ISS_075RB_BMOVIE4001_VIMS | 54 – 116 | 23.8 – 30.5 | 57.9 – 66.4 | 83.8 |
| 31 | 2008 AUG 02 | ISS_079RI_FMONITOR002_PRIME | 66 – 132 | 31.2 – 35.5 | 114.9 – 115.2 | 84.2 |
| 32 | 2008 AUG 06 | ISS_079RF_FRINGMRLF002_PRIME | 9 – 14 | 21.3 – 27.9 | 63.7 – 72.5 | 84.3 |
| 33 | 2008 AUG 14 | ISS_080RF_FMOVIE005_PRIME | 7 – 16 | 14.3 – 14.9 | 81.4 – 83.6 | 84.4 |
| 34 | 2008 AUG 16 | ISS_081RI_FMONITOR001_PRIME | 69 – 202 | 26.5 – 32.0 | 109.7 – 109.9 | 84.4 |
| 35 | 2008 AUG 20 | ISS_081RI_FMOVIE106_VIMS | 3 – 5 | 29.3 – 47.4 | 39.5 – 60.4 | 84.5 |
| 36 | 2008 AUG 28 | ISS_082RI_FMONITOR003_PRIME | 60 – 162 | 23.4 – 34.2 | 70.2 – 71.4 | 84.6 |
| 37 | 2008 AUG 30 | ISS_083RI_FMOVIE109_VIMS | 7 – 8 | 20.9 – 26.5 | 103.1 – 110.2 | 84.7 |
| 38 | 2008 SEP 08 | ISS_084RI_FMONITOR002_PRIME | 63 – 98 | 34.0 – 38.4 | 118.4 – 118.8 | 84.8 |
| 39 | 2008 SEP 15 | ISS_085RF_FMOVIE003_PRIME_1 | 5 – 6 | 41.4 – 47.8 | 126.3 – 132.5 | 84.9 |
| 40 | 2008 SEP 16 | ISS_085RF_FMOVIE003_PRIME_2 | 5 – 6 | 47.8 – 52.9 | 130.5 – 135.6 | 84.9 |
| 41 | 2008 SEP 30 | ISS_087RF_FMOVIE003_PRIME | 5 – 6 | 44.8 – 54.0 | 125.7 – 136.6 | 85.2 |
| 42 | 2008 OCT 14 | ISS_089RF_FMOVIE003_PRIME | 6 – 6 | 39.0 – 45.9 | 118.1 – 127.1 | 85.4 |
| 43 | 2008 OCT 29 | ISS_091RF_FMOVIE003_PRIME | 6 – 9 | 43.6 – 55.7 | 127.9 – 141.2 | 85.6 |
| 44 | 2008 NOV 02 | ISS_091RI_APOMOSL109_VIMS | 49 – 153 | 30.0 – 46.1 | 56.1 – 67.7 | 85.6 |
| 45 | 2008 NOV 21 | ISS_094RF_FMOVIE001_PRIME | 6 – 7 | 36.2 – 40.8 | 122.0 – 126.8 | 86.0 |
| 46 | 2008 DEC 10 | ISS_096RF_FMOVIE004_PRIME | 4 – 5 | 53.1 – 74.0 | 17.5 – 40.5 | 86.2 |
| 47 | 2008 DEC 23 | ISS_098RI_TMAPN30LP001_CIRS | 63 – 123 | 34.0 – 37.7 | 116.0 – 121.8 | 86.4 |
| 48 | 2009 JAN 11 | ISS_100RF_FMOVIE003_PRIME | 6 – 6 | 36.2 – 47.1 | 122.8 – 133.5 | 86.7 |
| 49 | 2009 JAN 17 | ISS_100RI_SUBMS20LP001_CIRS | 67 – 185 | 42.7 – 45.5 | 66.1 – 68.0 | 86.8 |
| 50 | 2009 FEB 05 | ISS_102RI_SPKFMLFLP001_PRIME | 66 – 163 | 36.5 – 41.9 | 59.9 – 66.3 | 87.1 |
| 51 | 2009 FEB 10 | ISS_103RF_FMOVIE003_PRIME | 6 – 6 | 69.8 – 82.1 | 150.1 – 155.4 | 87.2 |
| 52 | 2009 MAR 04 | ISS_105RF_FMOVIE003_PRIME | 7 – 7 | 33.6 – 34.9 | 110.9 – 114.3 | 87.5 |
| 53 | 2009 MAR 05 | ISS_105RI_TMAPN45LP001_CIRS | 68 – 193 | 40.5 – 49.0 | 128.0 – 135.8 | 87.6 |
| 54 | 2009 MAR 10 | ISS_105RI_TDIFS20HP001_CIRS | 48 – 103 | 151.8 – 157.0 | 64.6 – 71.5 | 87.6 |
| 55 | 2009 MAR 11 | ISS_105RF_FMOVIE002_PRIME | 5 – 6 | 98.9 – 111.9 | 23.3 – 25.5 | 87.7 |
| 56 | 2009 MAR 23 | ISS_106RF_FMOVIE002_PRIME | 5 – 6 | 99.9 – 112.3 | 23.3 – 25.6 | 87.8 |
| 57 | 2009 MAR 29 | ISS_107RF_FMOVIE002_PRIME | 7 – 7 | 54.9 – 61.1 | 139.8 – 144.0 | 87.9 |
| 58 | 2009 APR 10 | ISS_108RI_SPKMVLFLP001_PRIME | 78 – 237 | 35.2 – 39.9 | 68.5 – 73.7 | 88.1 |
| 59 | 2009 APR 16 | ISS_108RF_FMOVIE001_PRIME | 8 – 8 | 100.8 – 107.7 | 142.4 – 147.1 | 88.2 |
| 60 | 2009 APR 21 | ISS_109RI_TDIFS20HP001_CIRS | 66 – 125 | 144.2 – 148.7 | 56.0 – 60.6 | 88.3 |
| 61 | 2009 MAY 26 | ISS_111RF_FMOVIE002_PRIME | 5 – 24 | 34.3 – 38.2 | 93.1 – 103.1 | 88.8 |
| 62 | 2009 JUN 10 | ISS_112RF_FMOVIE002_PRIME | 4 – 6 | 17.1 – 21.5 | 97.8 – 109.6 | 89.1 |
| 63 | 2009 JUL 13 | ISS_114RF_FMOVIEEQX001_PRIME | 9 – 10 | 88.2 – 92.2 | 130.9 – 132.7 | 89.6 |
| 64 | 2009 JUL 30 | ISS_115RF_FMOVIEEQX001_PRIME | 10 – 11 | 98.9 – 101.8 | 117.5 – 118.9 | 89.8 |
| 65 | 2010 MAY 31 | ISS_132RI_FMOVIE001_VIMS | 7 – 12 | 120.0 – 125.6 | 79.5 – 80.1 | 85.5 |

| Obs # | Date | Cassini Observation ID | First Image | Last Image | # of Images | Coverage Percentage | # of ECs | # of MDCs |
|---|---|---|---|---|---|---|---|---|
| 1 | 2004 JUN 20 | ISS_000RI_SATSRCHAP001_PRIME | N1466448221_1 | N1466504861_1 | 119 | 100.0 | 32 | 0 |
| 2 | 2004 NOV 15 | ISS_00ARI_SPKMOVPER001_PRIME | N1479201492_1 | N1479254052_1 | 73 | 100.0 | 31 | 0 |
| 3 | 2005 APR 13 | ISS_006RI_LPHRLFMOV001_PRIME | N1492052646_1 | N1492102189_1 | 1320 | 96.8 | 23 | 8 |
| 4 | 2005 MAY 01 | ISS_007RI_LPHRLFMOV001_PRIME | N1493613276_1 | N1493662416_1 | 246 | 89.4 | 32 | 8 |
| 5 | 2005 MAY 02 | ISS_007RI_AZSCNLOPH001_PRIME | W1493706056_1 | W1493734145_1 | 28 | 98.6 | 30 | 8 |
| 6 | 2005 MAY 03 | ISS_007RI_HPMRDFMOV001_PRIME | N1493850077_1 | N1493892777_1 | 244 | 66.1 | 18 | 0 |
| 7 | 2006 SEP 28 | ISS_029RF_FMOVIE001_VIMS | N1538168640_1 | N1538218132_1 | 93 | 93.6 | 22 | 3 |
| 8 | 2006 SEP 30 | ISS_029RF_FMOVIE002_VIMS | N1538269441_1 | N1538300071_1 | 54 | 59.4 | 13 | 2 |
| 9 | 2006 OCT 16 | ISS_030RF_FMOVIE001_VIMS | N1539655570_1 | N1539683497_1 | 32 | 53.4 | 13 | 1 |
| 10 | 2006 OCT 31 | ISS_031RF_FMOVIE001_VIMS | N1541012989_1 | N1541062380_1 | 112 | 93.0 | 25 | 8 |
| 11 | 2006 NOV 12 | ISS_032RF_FMOVIE001_VIMS | N1542047155_1 | N1542096546_1 | 112 | 92.9 | 22 | 10 |
| 12 | 2006 NOV 25 | ISS_033RF_FMOVIE001_VIMS | N1543166702_1 | N1543216891_1 | 99 | 95.3 | 22 | 7 |
| 13 | 2006 DEC 23 | ISS_036RF_FMOVIE001_VIMS | N1545556618_1 | N1545613256_1 | 128 | 100.0 | 28 | 3 |
| 14 | 2007 JAN 05 | ISS_036RF_FMOVIE002_VIMS | N1546700688_5 | N1546748805_1 | 127 | 88.7 | 26 | 1 |
| 15 | 2007 FEB 10 | ISS_039RF_FMOVIE002_VIMS | N1549801218_1 | N1549851279_1 | 123 | 91.7 | 19 | 1 |
| 16 | 2007 FEB 27 | ISS_039RF_FMOVIE001_VIMS | N1551253524_1 | N1551310298_1 | 143 | 100.0 | 30 | 4 |
| 17 | 2007 MAR 17 | ISS_041RF_FMOVIE002_VIMS | N1552790437_1 | N1552850917_1 | 168 | 100.0 | 24 | 5 |
| 18 | 2007 MAR 31 | ISS_041RF_FMOVIE001_VIMS | N1554026927_1 | N1554072073_1 | 130 | 85.1 | 19 | 4 |
| 19 | 2007 APR 18 | ISS_043RF_FMOVIE001_VIMS | N1555557017_1 | N1555613413_1 | 94 | 100.0 | 31 | 3 |
| 20 | 2007 MAY 05 | ISS_044RF_FMOVIE001_VIMS | N1557020880_1 | N1557086720_1 | 175 | 100.0 | 31 | 2 |
| 21 | 2007 OCT 15 | ISS_051RI_LPMRDFMOV001_PRIME | N1571435192_1 | N1571475337_1 | 258 | 86.5 | 19 | 0 |
| 22 | 2007 DEC 31 | ISS_055RF_FMOVIE001_VIMS | N1577809417_1 | N1577857957_1 | 149 | 91.3 | 30 | 7 |
| 23 | 2008 JAN 07 | ISS_055RI_LPMRDFMOV001_PRIME | N1578386361_1 | N1578440131_1 | 190 | 100.0 | 34 | 7 |
| 24 | 2008 JAN 23 | ISS_057RF_FMOVIE001_VIMS | N1579790806_1 | N1579837831_1 | 131 | 89.5 | 25 | 5 |
| 25 | 2008 FEB 17 | ISS_059RF_FMOVIE001_VIMS | N1581944506_1 | N1581993408_1 | 116 | 92.2 | 28 | 8 |
| 26 | 2008 FEB 24 | ISS_059RF_FMOVIE002_VIMS | N1582549430_1 | N1582606673_1 | 132 | 100.0 | 28 | 8 |
| 27 | 2008 MAR 15 | ISS_061RI_LPMRDFMOV001_PRIME | N1584269462_1 | N1584298342_1 | 152 | 59.7 | 18 | 2 |
| 28 | 2008 JUN 14 | ISS_072RI_SPKHRLPDF001_PRIME | W1592114050_1 | W1592159350_1 | 76 | 94.0 | 24 | 2 |
| 29 | 2008 JUL 05 | ISS_075RF_FMOVIE002_VIMS | N1593913221_1 | N1593969867_1 | 111 | 98.9 | 28 | 2 |
| 30 | 2008 JUL 08 | ISS_075RB_BMOVIE4001_VIMS | W1594182967_1 | W1594205050_1 | 30 | 42.0 | 10 | 1 |
| 31 | 2008 AUG 02 | ISS_079RI_FMONITOR002_PRIME | W1596333808_1 | W1596335548_1 | 2 | 39.6 | 11 | 0 |
| 32 | 2008 AUG 06 | ISS_079RF_FRINGMRLF002_PRIME | N1596680431_1 | N1596713637_1 | 163 | 62.7 | 17 | 1 |
| 33 | 2008 AUG 14 | ISS_080RF_FMOVIE005_PRIME | N1597390145_1 | N1597402524_1 | 62 | 29.0 | 5 | 0 |
| 34 | 2008 AUG 16 | ISS_081RI_FMONITOR001_PRIME | W1597577017_1 | W1597578712_7 | 2 | 53.2 | 13 | 0 |
| 35 | 2008 AUG 20 | ISS_081RI_FMOVIE106_VIMS | N1597886079_1 | N1597933535_1 | 221 | 87.4 | 25 | 1 |
| 36 | 2008 AUG 28 | ISS_082RI_FMONITOR003_PRIME | W1598607164_1 | W1598612144_1 | 2 | 34.2 | 12 | 0 |
| 37 | 2008 AUG 30 | ISS_083RI_FMOVIE109_VIMS | N1598806665_1 | N1598853071_1 | 222 | 88.9 | 26 | 2 |
| 38 | 2008 SEP 08 | ISS_084RI_FMONITOR002_PRIME | W1599539571_1 | W1599541251_1 | 2 | 32.4 | 10 | 2 |
| 39 | 2008 SEP 15 | ISS_085RF_FMOVIE003_PRIME_1 | N1600213195_1 | N1600239816_1 | 123 | 49.4 | 13 | 2 |
| 40 | 2008 SEP 16 | ISS_085RF_FMOVIE003_PRIME_2 | N1600239816_1 | N1600258555_1 | 89 | 35.5 | 7 | 0 |
| 41 | 2008 SEP 30 | ISS_087RF_FMOVIE003_PRIME | N1601485634_1 | N1601526770_1 | 171 | 77.5 | 28 | 7 |
| 42 | 2008 OCT 14 | ISS_089RF_FMOVIE003_PRIME | N1602717403_1 | N1602760410_1 | 174 | 81.0 | 22 | 6 |
| 43 | 2008 OCT 29 | ISS_091RF_FMOVIE003_PRIME | N1604005372_1 | N1604050740_1 | 200 | 85.4 | 27 | 2 |
| 44 | 2008 NOV 02 | ISS_091RI_APOMOSL109_VIMS | W1604279522_1 | W1604299518_1 | 33 | 55.1 | 21 | 0 |
| 45 | 2008 NOV 21 | ISS_094RF_FMOVIE001_PRIME | N1605996366_1 | N1606021302_1 | 44 | 47.9 | 19 | 0 |
| 46 | 2008 DEC 10 | ISS_096RF_FMOVIE004_PRIME | N1607625633_1 | N1607670827_1 | 279 | 85.0 | 26 | 0 |
| 47 | 2008 DEC 23 | ISS_098RI_TMAPN30LP001_CIRS | W1608683935_1 | W1608705204_1 | 29 | 69.6 | 16 | 0 |
| 48 | 2009 JAN 11 | ISS_100RF_FMOVIE003_PRIME | N1610364098_1 | N1610404395_1 | 211 | 74.1 | 19 | 0 |
| 49 | 2009 JAN 17 | ISS_100RI_SUBMS20LP001_CIRS | W1610902092_1 | W1610943792_1 | 48 | 63.1 | 18 | 1 |
| 50 | 2009 FEB 05 | ISS_102RI_SPKFMLFLP001_PRIME | W1612545569_1 | W1612574369_1 | 9 | 61.6 | 20 | 4 |
| 51 | 2009 FEB 10 | ISS_103RF_FMOVIE003_PRIME | N1612969737_1 | N1613007123_1 | 213 | 63.7 | 18 | 3 |
| 52 | 2009 MAR 04 | ISS_105RF_FMOVIE003_PRIME | N1614850030_1 | N1614865561_1 | 65 | 30.4 | 12 | 1 |
| 53 | 2009 MAR 05 | ISS_105RI_TMAPN45LP001_CIRS | W1614936340_1 | W1614959136_1 | 32 | 32.3 | 12 | 1 |
| 54 | 2009 MAR 10 | ISS_105RI_TDIFS20HP001_CIRS | W1615342663_1 | W1615360723_1 | 36 | 47.5 | 13 | 1 |
| 55 | 2009 MAR 11 | ISS_105RF_FMOVIE002_PRIME | N1615465964_1 | N1615514239_1 | 210 | 78.4 | 24 | 4 |
| 56 | 2009 MAR 23 | ISS_106RF_FMOVIE002_PRIME | N1616500071_1 | N1616546465_1 | 209 | 75.1 | 19 | 4 |
| 57 | 2009 MAR 29 | ISS_107RF_FMOVIE002_PRIME | N1617039146_1 | N1617062017_1 | 126 | 41.5 | 9 | 2 |
| 58 | 2009 APR 10 | ISS_108RI_SPKMVLFLP001_PRIME | W1618050603_1 | W1618070583_1 | 19 | 60.4 | 22 | 3 |
| 59 | 2009 APR 16 | ISS_108RF_FMOVIE001_PRIME | N1618571707_1 | N1618607233_1 | 210 | 64.8 | 18 | 0 |
| 60 | 2009 APR 21 | ISS_109RI_TDIFS20HP001_CIRS | W1619011390_1 | W1619036946_1 | 24 | 70.9 | 17 | 0 |
| 61 | 2009 MAY 26 | ISS_111RF_FMOVIE002_PRIME | N1622022571_1 | N1622049830_1 | 199 | 53.4 | 16 | 0 |
| 62 | 2009 JUN 10 | ISS_112RF_FMOVIE002_PRIME | N1623328380_1 | N1623354366_1 | 115 | 50.1 | 14 | 0 |
| 63 | 2009 JUL 13 | ISS_114RF_FMOVIEEQX001_PRIME | N1626209041_1 | N1626252768_1 | 130 | 83.4 | 27 | 1 |
| 64 | 2009 JUL 30 | ISS_115RF_FMOVIEEQX001_PRIME | N1627609661_1 | N1627654945_1 | 147 | 87.4 | 23 | 1 |
| 65 | 2010 MAY 31 | ISS_132RI_FMOVIE001_VIMS | N1654040868_1 | N1654086464_1 | 90 | 82.2 | 25 | 0 |

**Table 2: Cassini multiply-detected clumps.** C19/2006 and C54/2009 are anomalous clumps discussed in the text and are given year suffixes to emphasize their unique status. First and Last ID # refer to the observations in Table 1. # of Obs is the number of observations in which the clump was detected. Longitude is in the co-rotating reference frame and is for the first observation of the clump. Minimum lifetime is the actually observed lifetime; maximum lifetime is the largest possible lifetime based on earlier and later observations where the clump was not observed. Angular width change is described in Section 4.3 and is only given for clumps seen for more than two weeks. Median angular width and brightness are the median of all observations of the clump; BN is baseline-normalized brightness, PN is phase-normalized brightness and is only presented for clumps where all observations have $|B_0| > 3°$.

| Clump Number | First ID # | Last ID # | # of Obs | Longitude at Epoch (°) | Min – Max Lifetime (days) | Relative Mean Motion (°/day) | Semimajor Axis (km) | Median Width (°) | Width Change (°/day) | Median BN Int Brt (°) | Median PN Int Brt (km °) | Media BN Peak Brt | Median PN Peak Brt (km) |
|---|---|---|---|---|---|---|---|---|---|---|---|---|---|
| C1 | 3 | 5 | 3 | 26.9 | 19 – 169 | 0.024 ± 0.010 | 140217.5 ± 1.6 | 5.8 | 0.01 | 0.668 | 0.222 | 0.222 | 0.074 |
| C2 | 3 | 5 | 3 | 38.5 | 19 – 169 | 0.010 ± 0.017 | 140219.7 ± 2.8 | 5.8 | 0.00 | 0.833 | 0.298 | 0.301 | 0.101 |
| C3 | 3 | 5 | 3 | 43.3 | 19 – 169 | 0.035 ± 0.010 | 140215.6 ± 1.6 | 5.8 | 0.09 | 1.043 | 0.412 | 0.387 | 0.130 |
| C4 | 3 | 5 | 3 | 218.1 | 19 – 169 | 0.114 ± 0.003 | 140202.9 ± 0.4 | 7.7 | 0.04 | 1.524 | 0.535 | 0.338 | 0.130 |
| C5 | 3 | 5 | 3 | 251.5 | 19 – 682 | 0.082 ± 0.005 | 140208.1 ± 0.7 | 4.8 | 0.01 | 0.931 | 0.372 | 0.357 | 0.130 |
| C6 | 3 | 5 | 3 | 263.8 | 19 – 682 | -0.025 ± 0.013 | 140225.4 ± 2.2 | 5.9 | 0.00 | 0.810 | 0.272 | 0.281 | 0.101 |
| C7 | 3 | 5 | 3 | 270.2 | 19 – 682 | 0.010 ± 0.001 | 140219.8 ± 0.1 | 4.6 | 0.03 | 0.788 | 0.351 | 0.273 | 0.122 |
| C8 | 3 | 5 | 3 | 348.8 | 19 – 682 | 0.068 ± 0.023 | 140210.4 ± 3.7 | 7.0 | –0.13 | 3.076 | 1.372 | 0.903 | 0.402 |
| C9 | 7 | 12 | 5 | 57.3 | 58 – 599 | 0.029 ± 0.009 | 140216.6 ± 1.4 | 5.7 | –0.03 | 0.688 | 0.315 | 0.195 | 0.091 |
| C10 | 7 | 13 | 6 | 136.8 | 86 – 612 | 0.023 ± 0.007 | 140217.6 ± 1.2 | 16.3 | 0.09 | 2.272 | 1.062 | 0.268 | 0.126 |
| C11 | 7 | 12 | 5 | 223.7 | 58 – 599 | 0.090 ± 0.005 | 140206.9 ± 0.9 | 6.2 | –0.03 | 1.237 | 0.572 | 0.343 | 0.160 |
| C12 | 10 | 11 | 2 | 70.3 | 12 – 57 | -0.214 ± 0.035 | 140255.7 ± 5.7 | 7.5 | N/A | 1.041 | 0.487 | 0.303 | 0.142 |
| C13 | 10 | 11 | 2 | 184.3 | 12 – 85 | 0.029 ± 0.035 | 140216.7 ± 5.7 | 8.2 | N/A | 0.793 | 0.371 | 0.190 | 0.089 |
| C14 | 10 | 11 | 2 | 195.6 | 12 – 86 | -0.040 ± 0.035 | 140227.7 ± 5.7 | 10.3 | N/A | 1.056 | 0.494 | 0.192 | 0.090 |
| C15 | 10 | 12 | 3 | 206.1 | 25 – 86 | -0.018 ± 0.009 | 140224.2 ± 1.4 | 13.6 | –0.03 | 1.112 | 0.520 | 0.159 | 0.075 |
| C16 | 10 | 12 | 3 | 283.2 | 25 – 69 | -0.046 ± 0.009 | 140228.7 ± 1.4 | 11.8 | –0.11 | 1.072 | 0.501 | 0.169 | 0.079 |
| C17 | 11 | 13 | 3 | 86.6 | 41 – 199 | 0.105 ± 0.019 | 140204.5 ± 3.0 | 6.8 | –0.10 | 0.287 | 0.136 | 0.084 | 0.044 |
| C18 | 11 | 12 | 2 | 107.3 | 13 – 85 | -0.085 ± 0.033 | 140235.0 ± 5.3 | 8.1 | N/A | 0.645 | 0.303 | 0.161 | 0.076 |
| C19/2006 | 13 | 20 | 8 | 217.1 | 133 – 327 | 0.575 ± 0.013 | 140129.0 ± 2.0 | 88.0 | 0.63 | 99.190 | 45.743 | 2.753 | 1.538 |
| C20 | 16 | 17 | 2 | 51.2 | 18 – 49 | -0.078 ± 0.024 | 140233.9 ± 3.8 | 12.4 | 0.19 | 3.488 | 1.765 | 0.531 | 0.268 |
| C21 | 16 | 20 | 5 | 85.8 | 67 – 360 | 0.127 ± 0.009 | 140200.9 ± 1.4 | 6.4 | –0.08 | 1.218 | 0.708 | 0.288 | 0.141 |
| C22 | 16 | 18 | 3 | 103.0 | 32 – 116 | -0.010 ± 0.027 | 140223.0 ± 4.4 | 18.4 | 0.79 | 5.741 | 2.980 | 0.909 | 0.472 |
| C23 | 17 | 19 | 3 | 304.6 | 32 – 67 | -0.175 ± 0.014 | 140249.4 ± 2.2 | 6.6 | –0.03 | 2.290 | 1.186 | 0.674 | 0.349 |
| C24 | 22 | 24 | 3 | 34.0 | 23 – 129 | 0.021 ± 0.019 | 140217.9 ± 3.0 | 5.1 | 0.12 | 0.609 | 0.332 | 0.212 | 0.113 |
| C25 | 22 | 24 | 3 | 43.7 | 23 – 122 | 0.009 ± 0.017 | 140219.9 ± 2.8 | 9.8 | 0.06 | 0.641 | 0.396 | 0.153 | 0.075 |
| C26 | 22 | 23 | 2 | 51.7 | 7 – 97 | 0.101 ± 0.063 | 140205.2 ± 10.1 | 7.3 | N/A | 0.575 | 0.315 | 0.167 | 0.093 |
| C27 | 22 | 23 | 2 | 123.0 | 7 – 263 | -0.247 ± 0.063 | 140261.1 ± 10.1 | 9.5 | N/A | 0.844 | 0.470 | 0.197 | 0.107 |
| C28 | 22 | 23 | 2 | 221.0 | 7 – 97 | 0.189 ± 0.063 | 140190.9 ± 10.1 | 5.4 | N/A | 1.532 | 0.837 | 0.523 | 0.283 |
| C29 | 22 | 24 | 3 | 259.2 | 23 – 122 | 0.064 ± 0.009 | 140211.1 ± 1.5 | 6.9 | 0.08 | 1.184 | 0.605 | 0.331 | 0.161 |
| C30 | 22 | 26 | 4 | 319.3 | 55 – 148 | -0.102 ± 0.006 | 140237.7 ± 1.0 | 5.7 | 0.03 | 1.269 | 0.630 | 0.349 | 0.185 |
| C31 | 24 | 26 | 3 | 61.0 | 32 – 180 | -0.006 ± 0.004 | 140222.3 ± 0.6 | 5.5 | 0.01 | 0.350 | 0.169 | 0.145 | 0.070 |
| C32 | 24 | 27 | 4 | 248.4 | 52 – 159 | -0.058 ± 0.007 | 140230.6 ± 1.1 | 6.1 | –0.04 | 0.928 | 0.458 | 0.275 | 0.136 |
| C33 | 25 | 26 | 2 | 56.4 | 7 – 143 | 0.088 ± 0.060 | 140207.2 ± 9.6 | 4.7 | N/A | 0.294 | 0.157 | 0.122 | 0.065 |
| C34 | 25 | 26 | 2 | 151.8 | 7 – 52 | -0.042 ± 0.060 | 140228.0 ± 9.6 | 8.2 | N/A | 1.101 | 0.601 | 0.239 | 0.129 |
| C35 | 25 | 26 | 2 | 279.8 | 7 – 52 | -0.006 ± 0.060 | 140222.3 ± 9.6 | 26.7 | N/A | 11.416 | 6.142 | 1.145 | 0.608 |
| C36 | 25 | 26 | 2 | 324.2 | 7 – 68 | 0.019 ± 0.060 | 140218.2 ± 9.6 | 4.3 | N/A | 0.564 | 0.296 | 0.223 | 0.118 |
| C37 | 25 | 27 | 3 | 335.3 | 27 – 159 | -0.076 ± 0.003 | 140233.6 ± 0.4 | 8.7 | –0.15 | 1.088 | 0.561 | 0.237 | 0.121 |
| C38 | 28 | 30 | 3 | 87.5 | 24 – 159 | -0.094 ± 0.028 | 140236.4 ± 4.5 | 6.5 | 0.07 | 1.480 | 0.680 | 0.479 | 0.205 |
| C39 | 28 | 29 | 2 | 268.7 | 21 – 139 | 0.002 ± 0.020 | 140221.0 ± 3.3 | 11.6 | –0.05 | 2.443 | 1.078 | 0.397 | 0.175 |
| C40 | 32 | 37 | 3 | 180.6 | 25 – 73 | -0.038 ± 0.006 | 140227.5 ± 1.0 | 8.3 | –0.20 | 1.061 | 0.480 | 0.243 | 0.107 |
| C41 | 37 | 39 | 3 | 95.5 | 16 – 48 | 0.011 ± 0.005 | 140219.6 ± 0.8 | 11.2 | –0.01 | 1.564 | 0.646 | 0.319 | 0.120 |
| C42 | 38 | 41 | 3 | 104.9 | 23 – 60 | 0.044 ± 0.043 | 140214.2 ± 6.9 | 8.7 | –0.15 | 1.292 | 0.464 | 0.411 | 0.147 |
| C43 | 41 | 42 | 2 | 19.4 | 14 – 94 | 0.066 ± 0.030 | 140210.8 ± 4.8 | 8.1 | 0.12 | 1.689 | 0.632 | 0.373 | 0.139 |
| C44 | 41 | 42 | 2 | 32.8 | 14 – 60 | -0.058 ± 0.030 | 140230.6 ± 4.8 | 9.0 | 0.11 | 2.870 | 1.078 | 0.679 | 0.254 |
| C45 | 41 | 42 | 2 | 49.8 | 14 – 60 | -0.034 ± 0.030 | 140226.8 ± 4.8 | 11.3 | 0.04 | 1.406 | 0.523 | 0.251 | 0.094 |
| C46 | 41 | 43 | 3 | 69.8 | 29 – 75 | 0.010 ± 0.015 | 140219.8 ± 2.4 | 4.7 | –0.10 | 0.566 | 0.218 | 0.214 | 0.083 |
| C47 | 41 | 43 | 3 | 305.7 | 29 – 63 | -0.009 ± 0.002 | 140222.8 ± 0.3 | 9.0 | –0.03 | 1.554 | 0.599 | 0.383 | 0.148 |
| C48 | 41 | 42 | 2 | 316.1 | 14 – 60 | 0.166 ± 0.030 | 140194.6 ± 4.8 | 11.8 | 0.12 | 3.017 | 1.127 | 0.514 | 0.191 |
| C49 | 49 | 50 | 2 | 103.9 | 19 – 53 | 0.017 ± 0.023 | 140218.6 ± 3.6 | 6.8 | –0.04 | 1.465 | N/A | 0.415 | N/A |
| C50 | 50 | 51 | 2 | 214.1 | 5 – 53 | 0.242 ± 0.086 | 140182.4 ± 13.8 | 5.2 | N/A | 0.847 | N/A | 0.284 | N/A |
| C51 | 50 | 51 | 2 | 232.4 | 5 – 53 | -0.068 ± 0.086 | 140232.2 ± 13.8 | 4.9 | N/A | 0.632 | N/A | 0.222 | N/A |
| C52 | 50 | 51 | 2 | 256.5 | 5 – 51 | 0.057 ± 0.086 | 140212.2 ± 13.8 | 5.3 | N/A | 0.614 | N/A | 0.231 | N/A |
| C53 | 52 | 55 | 3 | 41.7 | 8 – 65 | 0.001 ± 0.002 | 140221.1 ± 0.2 | 4.8 | N/A | 0.671 | N/A | 0.223 | N/A |
| C54/2009 | 54 | 58 | 4 | 317.3 | 31 – 65 | 0.418 ± 0.024 | 140154.2 ± 3.8 | 19.2 | 0.77 | 9.530 | N/A | 1.251 | N/A |
| C55 | 55 | 56 | 2 | 110.6 | 12 – 42 | 0.167 ± 0.035 | 140194.5 ± 5.7 | 8.4 | N/A | 3.248 | N/A | 0.879 | N/A |
| C56 | 55 | 58 | 4 | 293.6 | 30 – 43 | 0.003 ± 0.023 | 140220.8 ± 3.7 | 5.5 | 0.11 | 0.609 | N/A | 0.184 | N/A |
| C57 | 56 | 58 | 3 | 288.8 | 18 – 76 | -0.094 ± 0.024 | 140236.5 ± 3.9 | 5.0 | –0.04 | 0.521 | N/A | 0.207 | N/A |
| C58 | 63 | 64 | 2 | 121.0 | 16 – 406 | 0.032 ± 0.026 | 140216.2 ± 4.2 | 6.3 | –0.10 | 1.230 | N/A | 0.379 | N/A |

**Table 3: Physical characteristics of ECs seen by Voyager and Cassini. The Kolmogorov-Smirnov test is used to show how similar the Voyager and Cassini distributions are; large *D* and small *p* imply that the distributions are different.**

| Characteristic | Voyager | | Cassini | | Kolmogorov-Smirnov |
| --- | --- | --- | --- | --- | --- |
| | Range | Mean | Range | Mean | Test |
| Angular width (°) | | | | | |
|   C19/2006 | | | 42.0 – 124.0 | 86.5 ± 31.0 | |
|   C54/2009 | | | 6.5 – 30.5 | 19.0 ± 10.8 | |
|   All other ECs | 4.5 – 37.5 | 12.5 ± 7.5 | 3.5 – 40.0 | 11.5 ± 7.4 | $D = 0.10, p = 0.43$ |
| | | | | | |
| Phase-normalized integrated brightness (km°) | | | | | |
|   C19/2006 | | | 24.29 – 69.73 | 48.36 ± 14.08 | |
|   C54/2009 | | | N/A | N/A | |
|   All other ECs | 0.06 – 7.45 | 0.95 ± 1.41 | 0.04 – 8.72 | 0.71 ± 0.91 | $D = 0.08, p = 0.70$ |
|     Brightness/angular width slope (km) | | 0.115 ± 0.016 | | 0.100 ± 0.002 | |
| | | | | | |
| Phase-normalized peak brightness (km) | | | | | |
|   C19/2006 | | | 0.94 to 2.48 | 1.56 ± 0.44 | |
|   C54/2009 | | | N/A | N/A | |
|   All other ECs | 0.03 – 0.94 | 0.13 ± 0.16 | 0.02 – 0.83 | 0.10 ± 0.08 | $D = 0.09, p = 0.60$ |
|     Brightness/angular width slope (km/°) | | 0.0082 ± 0.0022 | | 0.0049 ± 0.0003 | |
| | | | | | |
| Baseline-normalized integrated brightness (°) | | | | | |
|   C19/2006 | | | 46.54 – 134.40 | 94.82 ± 28.23 | |
|   C54/2009 | | | 2.00 – 19.02 | 10.00 ± 7.46 | |
|   All other ECs | 0.35 – 41.30 | 5.15 ± 7.37 | 0.07 – 15.63 | 1.60 ± 1.99 | $D = 0.37, p = 0.00$ |
|     Brightness/angular width slope | | 0.586 ± 0.087 | | 0.218 ± 0.005 | |
| | | | | | |
| Baseline-normalized peak brightness | | | | | |
|   C19/2006 | | | 1.41 – 6.74 | 3.22 ± 1.51 | |
|   C54/2009 | | | 0.45 – 1.41 | 1.05 ± 0.40 | |
|   All other ECs | 0.14 – 5.17 | 0.73 ± 0.90 | 0.04 – 1.88 | 0.23 ± 0.18 | $D = 0.50, p = 0.00$ |
|     Brightness/angular width slope (deg$^{-1}$) | | 0.0402 ± 0.0126 | | 0.0106 ± 0.0007 | |